\documentclass[preprint,11pt]{aastex}
\usepackage{graphicx}
\usepackage{amssymb}
\usepackage{hyperref}
\usepackage{listings}
\usepackage{multicol}
\usepackage{epstopdf}

\usepackage[usenames,dvipsnames]{color}
\usepackage{rotating}

\usepackage[section]{placeins}

\begin{document}

\title{SPIDER - V. Measuring Systematic Effects in Early-Type Galaxy Stellar Masses from
Photometric SED Fitting}

\author{R. Swindle\footnote{swindle@ifa.hawaii.edu. High-res version at \url{http://www.ifa.hawaii.edu/users/swindle/SPIDER/PaperV.pdf}.}~~and R. R. Gal}
\affil{Institute for Astronomy, University of Hawai${\prime}$i}
\author{F. La Barbera}
\affil{INAF -- Osservatorio Astronomico di Capodimonte, Napoli, Italy}
\author{R.R. de Carvalho}
\affil{Instituto Nacional de Pesquisas Espaciais/MCT, S. J. dos Campos, Brazil}

\begin{abstract}

We present robust statistical estimates of the accuracy of early-type galaxy stellar masses derived from spectral energy distribution (SED) fitting as functions of various empirical and theoretical assumptions. Using large samples consisting of 40,000 galaxies from the Sloan Digital Sky Survey, of which 5,000 are also in the UKIRT Infrared Deep Sky Survey, with spectroscopic redshifts in the range 0.05 $\leq$ z $\leq$ 0.095, we test the reliability of some commonly used stellar population models and extinction laws for computing stellar masses. Spectroscopic ages (t), metallicities (Z), and extinctions (A) are also computed from fits to SDSS spectra using various population models. These constraints are used in additional tests to estimate the systematic errors in the stellar masses derived from SED fitting, where t, Z, and A are typically left as free parameters. We find reasonable agreement in mass estimates among stellar population models, with variation of the IMF and extinction law yielding systematic biases on the mass of nearly a factor of 2, in agreement with other studies. Removing the near-infrared bands changes the statistical bias in mass by only 0.06 dex, adding uncertainties of 0.1 dex at the 95\% CL. In contrast, we find that removing an ultraviolet band is more critical, introducing 2? uncertainties of 0.15 dex. Finally, we find that stellar masses are less affected by absence of metallicity and/or dust extinction knowledge. However, there is a definite systematic offset in the mass estimate when the stellar population age is unknown, up to a factor of 2.5 for very old (12 Gyr) stellar populations. We present the stellar masses for our sample, corrected for the measured systematic biases due to photometrically determined ages, finding that age errors produce lower stellar masses by 0.15 dex, with errors of 0.02 dex at the 95\% CL for the median stellar age subsample.

\end{abstract}

\keywords{galaxies: elliptical, ETG, stellar mass -- methods: SED fitting}

\section{Introduction}

Our understanding of the formation and evolution of early-type galaxies (ETGs)
represents a key ingredient in models of galaxy formation. Here, we consider ETGs to be
bulge dominated galaxies with passive spectra in their central
regions. Observationally, they are characterized by elliptical isophotes, generally
redder colors, and a sharp 4000\AA~break, corresponding to an accumulation of
absorption lines of mainly ionized metals, reflected in their rest frame UV/optical
colors. This feature is typical of old stellar populations with little to no
ongoing star formation.

At higher redshift, most ETGs can only be studied in integrated light, and
interpretation of their photometric and spectroscopic properties requires
population synthesis models. These single stellar population (SSP) models usually consist of stars born at the
same time with equal initial element compositions, evolved using the isochrone
synthesis technique (see \citealp{CB91} for a review), where stars of different
masses follow different evolutionary tracks. SSP models can be combined to produce arbitrary star-formation histories, although most studies restrict themselves to histories with exponentially declining star-formation. Such tau-models are parametrized by the e-folding time of this decline, $\tau$. The spectral energy distributions (SEDs) from a set of models with various parameters (e.g. initial mass function (IMF), star-formation history, age, metallicity, and
extinction) is compared to the photometric or spectroscopic observations to derive a best-fit template. A fundamental parameter derived from such SED fitting is the overall normalization of the model relative to the observations,  which gives the galaxy stellar mass content. Indeed, a measurement
of the stellar mass of an ETG is involved in many useful scaling relations, such
as the \emph{size-mass} relation \citep{size-mass} and \emph{downsizing}
\citep{downsizing}, where the evolutionary history of ETGs is seen to follow
different time scales as a function of their stellar mass content. Stellar mass
assembly in galaxies is also used in tests of hierarchical models, such as the 
evolution of the number density and size of both early and late-type galaxies as a function of redshift \citep{SDSS, GOODS, COSMOS}.
Furthermore, upcoming surveys (e.g. PanSTARRS, LSST, DES) will provide only
photometry, so it is crucial to understand how to obtain reliable stellar masses from
SED fitting techniques.

The degeneracies among the multiple model parameters which are required to reproduce the observed SEDs of the galaxies,  along with the available photometric bandpasses, determine the uncertainty of the mass estimates. For example, models and observations show that the rest-frame near-infrared (NIR) galaxy flux correlates well with stellar mass \citep{1998MNRAS.297L..23K, downsizing} due to weak contributions from hot, young stars and dust extinction at these wavelengths. But both different model predictions and/or absence of NIR data in fitting the SED can result in stellar masses which differ by a factor of 2, for both low and high redshift samples \citep{2006ApJ...652...97V}. Of particular interest in this study, the effects of unknown stellar population age, metallicity, and extinction on the stellar masses derived for elliptical galaxies have not been quantified. It is already well known that the age-metallicity degeneracy \citep{Worthey} cannot easily be broken with broadband colors alone.

All of these factors either depend on or affect the resulting evolution of the spectral energy distribution, models of which have been generated by, e.g. \cite{BC03}, \cite{Maraston05}, and \cite{PEGASE2}. Many phases in stellar evolution are still not well understood, but one key ingredient in these models, the thermally-pulsating asymptotic giant branch (TP-AGB) phase, is particularly influential for determining the galaxy stellar mass for a range of dominant stellar population ages. Light from these stars largely influences the integrated brightness of the NIR continuum, the effects of which were recently highlighted by \cite{Maraston06} and \cite{TP-AGB}, who found systematic differences of factors of $\sim$2 in mass and age over a large range in redshift. This example has shown us the effects of one sensitive parameter in the SED fitting process and demonstrates the need to further test the dependence of the stellar mass estimate on model parameters and the use of common ground-based observation filters.

Several authors \citep[e.g.][]{2009ApJ...699..486C, IMF_ETGs, 2011arXiv1103.0259R} have studied
some of these effects on the photometrically derived stellar mass, quantifying systematic offsets between masses calculated against different model parameters. However, they all suffer from a lack of spectroscopic data which can provide independent constraints on many of the galaxy properties.
\cite{2009MNRAS.394..774L} used a synthetic catalog of intermediate
redshift ($1 \leq z \leq 2$) ETGs to test the dependence of stellar
population models and assumed model parameters on stellar mass. However, they compare a synthetic
catalog to only 125 galaxies observed in the GOODS field, using a different SED
fitting code. \cite{COSMOS} and \cite{2010MNRAS.404.2087B} test the accuracy of
large surveys ($\sim$10,000 and $\sim$200,000 galaxies, respectively) at
estimating photometric stellar masses compared to those from available
spectroscopy. However, these projects select mostly galaxies at high redshift
and neglect the possible combined effects of assuming various free parameters.

This paper is part of a series examining the global and internal properties of ETGs in
the nearby Universe, combining optical and NIR photometry with spectroscopic data.
The Spheroids Panchromatic Investigation in Different Environmental Regions (SPIDER) project
is described in \citet[][hereafter Paper I]{PaperI}. The second and third papers of the 
series present a thorough analysis of the optical+NIR scaling relations of ETGs 
\citep{PaperII, PaperIII}, while in Paper IV \citep{PaperIV} we have analyzed the
optical+NIR internal color gradients of ETGs.
Here we present an extensive comparison of stellar mass estimates obtained for 
low redshift ETGs by varying several observational and model parameters 
(presented individually in Section \ref{sec_results}). The paper is organized as follows. 
Section \ref{sec_sample}
describes the galaxies, comprised of a carefully selected sample of high S/N,
low redshift ETGs. In Section~\ref{sec_models} we describe the different stellar 
population models used to compute model galaxy SEDs. 
In Section~\ref{sec_technique} we provide an overview of the general technique applied to fit 
the theoretical SEDs to the observed colors and the SDSS galaxy spectra.
The reliability of these fits is assessed in Section~\ref{sec_reliability}, by comparing 
photometric and spectroscopic fitting results. Our method for 
determining stellar masses is described in Section~\ref{sec_method}, while the results 
of the various model comparisons are discussed in Section \ref{sec_results}.
 Throughout the paper, we adopt a cosmology with H$_{0} = 70$
km s$^{-1}$ Mpc$^{-1}$, $\Omega_{m} = 0.3$, and $\Omega_{\Lambda} = 0.7$.

\section{The SPIDER sample}
\label{sec_sample}

\subsection{SDSS $ugriz$ sample of ETGs}
The sample of galaxies is selected from the Sloan Digital Sky Survey (SDSS) DR6
in the redshift range 0.05 to 0.095 and with $M_{r} < -20$, where $M_{r}$ is the
k-corrected SDSS Petrosian magnitude in the \emph{r}-band. The k-correction is estimated using the software \emph{kcorrect} \citep{2003ApJ...594..186B}, through a rest frame \emph{r}-band filter blue-shifted by a factor $(1+z_{0})$, where $z_{0}=0.0725$, the median redshift of the ETG sample (see Paper I, Section 3.1). The lower redshift limit is chosen to minimize the aperture bias
\citep{2003ApJ...584..210G}, while the upper redshift limit guarantees
a high level of completeness (according to \citealp{2006A&A...460..673S}) and allows us to define a volume-complete sample of \emph{bright}
early-type systems. ETGs follow two different trends in the size-luminosity
diagram \citep{1992MNRAS.259..323C, 2003AJ....125.2936G}. The separation between
these two families of \emph{bright} and \emph{ordinary} ellipticals occurs at an
absolute \emph{B}-band magnitude of $-$19, corresponding to the magnitude limit
of $M_{r} \sim -20$ adopted for this selection. At the upper redshift limit of
$z = 0.095$, the magnitude cut of $-20$ also corresponds approximately to the
magnitude limit where the SDSS spectroscopy is complete (i.e. Petrosian
magnitude of $m_{r} \sim 17.8$), making the sample volume limited. Following \cite{2003AJ....125.1849B}, we define ETGs using the SDSS spectroscopic parameter \emph{eClass}, which indicates the spectral type of a galaxy on the basis of a principal component analysis, and the SDSS photometric parameter \emph{fracDev$_{r}$}, which measures the fraction of galaxy light that is better fitted by a de Vaucouleurs (rather than an exponential) law. In this contribution, ETGs are those systems with $eClass < 0$ and $fracDev_{r} > 0.8$. The SDSS selection criteria and completeness of the ETG sample, part of the SPIDER project, are further detailed in Paper I.

\subsection{UKIDSS $YJHK$ photometry}
\label{sec_ukidss}
The SPIDER sample consists of 39,993 ETGs, with available \emph{ugriz}
photometry and spectroscopy. This SDSS sample is then matched to the UKIRT
Infrared Deep Sky Survey (UKIDSS) Large Area Survey DR4. The UKIDSS-LAS DR4
provides NIR photometry in the $YJHK$-bands over $\sim$1000 square degrees on
the sky, with significant overlap with SDSS. The $YHK$-band data have a pixel
scale of $0.4^{\prime\prime}$ pixel$^{-1}$, matching almost exactly the resolution of
the SDSS frames ($0.396^{\prime\prime}$ pixel$^{-1}$). $J$-band observations are
carried out with a resolution of $0.4^{\prime\prime}$ pixel$^{-1}$, and then
interleaved to a subpixel grid, resulting in stacked frames with
a resolution of $0.2^{\prime\prime}$ pixel$^{-1}$. The $YJHK$-stacked images
(multiframes) have average depths of 20.2, 19.6, 18.8, and 18.2 magnitudes (in the Vega system),
respectively. For each ETG in the SDSS sample, we searched for the nearest
UKIDSS detection within a radius of $1^{\prime\prime}$, considering only
UKIDSS frames with good quality flags ($ppErrBits < 16$). Of these
galaxies, 5,080 objects also have available photometry in the \emph{YJHK}
wavebands from UKIDSS. Hereafter, we refer to these samples as the complete and
optical+NIR samples of ETGs, respectively. 

\begin{figure}[h]
\begin{center}

\includegraphics[width=6.5in]{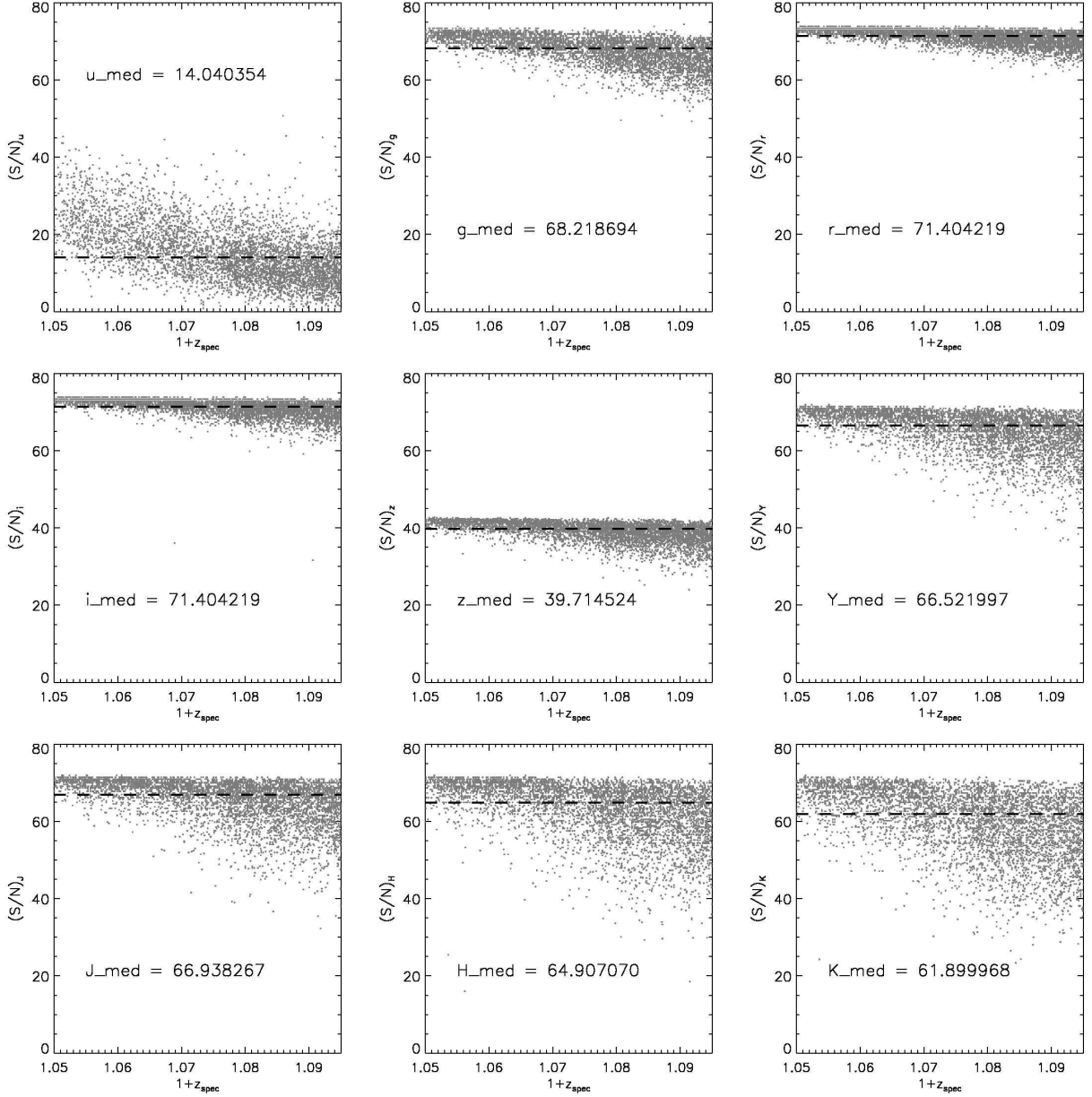}

\end{center}
\caption[XXX]
{\label{fig_SNR}
Signal-to-noise ratios (S/N) for the optical+NIR sample. The S/N in the
\emph{u}-band is too low to allow the fitting of galaxy images with Sersic models
(as in previous papers in this series; see Section \ref{sec_ukidss}), so SExtractor was used to 
obtain the magnitudes and corresponding errors. S/N in all of the other 
bands is remarkably high. $Dashed$ lines represent the median S/N in each bandpass.
}
\end{figure}

As detailed in Paper I, all \emph{grizYJHK} frames have been homogeneously processed with 2DPHOT \citep{2DPHOT}, an automated software environment that performs several tasks, such as catalog extraction (using SExtractor), star/galaxy separation, and galaxy surface photometry. For each ETG, magnitudes are measured within the same aperture in all $ugrizYJHK$ wavebands. Unless otherwise stated, in the present work we use the $ugrizYJHK$ magnitudes measured within an adaptive circular aperture of $3 \times r_{K,i}$, where $r_{K,i}$ is the Kron radius in the $i$-band. We use $i$-band Kron radii because of the larger S/N ratio of $i$ relative to $zYJHK$ frames (see Section~\ref{fig_SNR}), and its lower sensitivity to young stellar populations in a galaxy  relative to $ugr$. All magnitudes herein are in the AB system. For more than 95\% of all ETGs in the SPIDER sample, the $3 \times r_{K,i}$ aperture is at least three times larger than the seeing FWHM of the $grizYJHK$ frames, making the Kron magnitudes essentially independent of the seeing variation from  $g$ through $K$ (see Paper I). We also note that the first four papers in this series do not use $u$-band photometry, as the S/N of SDSS $u$-band frames is too low to measure reliable surface photometry. For the present study, we have processed the $u$-band images with SExtractor \citep{SExtractor}, using the same setup as for the $grizYJHK$ wavebands, now resulting in 5,068 objects in the optical+NIR sample. All magnitudes have been corrected for Galactic extinction, as detailed in Paper I.

\subsection{SDSS spectroscopy}
SDSS spectra in the range $3800-9200$\AA~are analyzed with the spectral fitting
code {\it STARLIGHT} \citep{STARLIGHT} to determine galaxy ages, metallicities, and
interstellar extinctions, among other parameters. {\it STARLIGHT} finds the combination
of SSP models that, normalized and broadened with a given sigma, best matches the observed
spectrum. To this effect, objects are de-redshifted and
corrected for foreground extinction, following the recipes described in Paper I. For the galaxy
spectra in the SPIDER sample,  the median value of the resolution
varies from $\sim$2.8\AA~(FWHM) in the blue (4000\AA) up to $\sim$3.7\AA~(FWHM) in the red (8000\AA). The theoretical spectra used for comparison
(and described below in Section \ref{sec_models}) have similar resolution 
to that of SDSS spectra across the whole wavelength range from $\sim$3000\AA~to 
$\sim$9000\AA~and are therefore not resampled for this study.

\section{Stellar population models}
\label{sec_models}
We begin by constructing a library of model spectra (SEDs), generated through different 
stellar population synthesis techniques that encompass a variety of stellar evolutionary
tracks. The models span a wide range in age, metallicity, extinction, and IMF.
In the following, we describe the SEDs used to fit the broadband colors  as well as
the spectra of ETGs.

\subsection{SEDs for fitting broadband colors}
\label{sec_bc03}

We use stellar population synthesis models to convert galaxy luminosity into
stellar mass \citep[e.g.][]{SDSS, 2004A&A...424...23F}. The stellar mass is derived from the
factor needed to rescale the spectrum from the best fit theoretical stellar
population (normalized at 1 $\mathcal{M}_{\sun}$) to the intrinsic (observed)
luminosities. The models we use to fit galaxy colors are described below.

\subsubsection{\cite{BC03}}

Among the variety of stellar evolution libraries provided by this code (hereafter BC03), we use
models from the  Padova 1994
library, which provide a median spectral resolution $R=2000$ over the
spectral range from 3200 to 9500\AA~(STELIB), and $R=300$ outside this range
(BaSeL 3.1; see BC03 for details). Models of stars with $T>50,000$ K are taken
from \cite{2003A&A...403..709R}, and the spectral models of the
thermally-pulsating asymptotic giant branch (TP-AGB) phase are based on
\cite{1993ApJ...413..641V}. We utilize models with four different metallicities:
$0.2Z_{\sun}, 0.4Z_{\sun}, Z_{\sun}, \mathrm{and}~2.5Z_{\sun}$. Within this
code, we also consider both \cite{Salpeter} and \cite{Chabrier} IMFs, with mass limits of $0.1M_{\sun} < M < 100M_{\sun}$. Unless otherwise noted, we
assume a Chabrier IMF due to its theoretical motivation. The star-formation
histories include exponentially declining $SFR \propto e^{-t/\tau}$ with the values of  $\tau$
given below, no gas recycling, and a $t=20$ Gyr cutoff time for star-formation.
This ensures that no models have their SF cut off before an e-folding time.
Table \ref{tab_ssp} provides a summary of the stellar population model parameters, where the last column provides stellar ages used in comparing {\it LePhare} and {\it STARLIGHT} results
(discussed in Section \ref{sec_reliability}).

\begin{table}[htdp]
\caption{Stellar population model parameters (BC03/CB10).}
\begin{center}
\begin{tabular}{cccc}

\hline
\hline
$\tau$ & & & $t_{SSP}$ \\
(Gyr) & $E(B-V)$ & $Z/Z_{\sun}$ & (Gyr) \\
\hline
0.1 & 0 & 0.2 & 0.5, 1.0, 2.0, \\
0.3 & 0.1 & 0.4 & 2.2, 2.5, 2.75, \\
1 & 0.2 & 1.0 & 3.25, 3.5, 4.0, \\
2 & 0.3 & 2.5 & 4.5, 5.0, 5.5, \\
3 & 0.4 & & 6.25, 7.0, 8.0, \\
5 & 0.5 & & 9.0, 10.0, 11.25 \\
10 & & & 12.5 \\
15 \\
30 \\
\hline
\end{tabular}
\end{center}
\label{tab_ssp}
\end{table}

The SEDs were generated for a grid of 64 ages in the range 0.8$-$14.2 Gyr. 
Since the mixing of dust and stars in (early-type) galaxies is far from
well understood, for the purposes of the present work we describe dust extinction
with a simplified approach, where attenuation is applied to the templates using the
\cite{Cardelli} law, with $E(B-V)$ in the range 0$-$0.5. The reddening $E(B-V)$ is limited to 0.5
magnitudes to avoid incorrect fitting of observationally red galaxies as highly
reddened blue galaxies; such objects should be absent from our sample. 
\cite{COSMOS} show that the stellar mass measurements are sensitive to the
extinction law, with absolute median differences of 0.14 and 0.27 dex for two
spectroscopic samples of high redshift galaxies when comparing the \cite{Calzetti} and
\citet[CF2000]{CF2000} extinction laws. We present similar results in Section
\ref{sec_ext}. Unless otherwise noted, we use the Cardelli extinction law throughout
this paper. We note that most studies using SED fitting to determine stellar masses have
used the Calzetti extinction law. However, this law is theoretically motivated for strong
starbursts. While this may be reasonable for field surveys, where the majority of galaxies
are likely to be star-forming, it is inappropriate for our sample, where we expect ETGs to
be dominated by older stellar populations. Therefore, we utilize the Cardelli relation as
our baseline.

\subsubsection{Charlot \& Bruzual (2010)}
An updated version of this code, which includes the new prescription of
\cite{2007A&A...469..239M} for the TP-AGB evolution of low and intermediate-mass
stars, is described in \cite{TP-AGB} and is used in previous papers in this series
(Papers II, III, IV). We consider its preliminary
version in the following comparisons, referring to it as the CB10 code (Charlot
\& Bruzual, private communication). These templates use the STELIB library over
the entire wavelength range covered by the  $ugrizYJHK$ bands. We use the same stellar
population parameters as BC03, described above and listed in Table
\ref{tab_ssp}.

\subsubsection{PEGASE.2: \cite{PEGASE2}}

The code by \cite{PEGASE2} is based on the same tracks and the same stellar
spectral library as BC03. The difference between the spectrophotometric models
is given by different stellar spectra assumptions for the hottest stars
($T>50,000 K$) and a different prescription for the TP-AGB phase, both affecting
the results at young ages ($\leq 2$ Gyr). Indeed, spectra of hot stars are taken
from \cite{1987MNRAS.228..759C}, while the models of the TP-AGB phase are based
on the prescriptions of \cite{1993A&A...267..410G}. For comparison with the
results obtained with the BC03/CB10 codes, we have used models with
exponentially declining SFR with time scales $\tau=0.1, 0.3, 1.0, 2.0, 3.0, 5.0,
10.0, \mathrm{and}~15.0$ Gyr and a \cite{Scalo86} IMF. Solar metallicity is
assumed for each template, based on the readily available models within
{\it LePhare}. The models are set up with no infall  (i.e. all the gas available to
form stars is assumed to be in place at time $t=0$) and no galactic wind, and the default value of 0.05 has been assumed for the parameter
representing the fraction of close binary systems. Because this library is
available in the public version of {\it LePhare},  we choose
to compare its results here.

The SEDs were generated for a grid of 39 ages (in the range 0.8$-$14.2 Gyr). 
Dust extinction was applied to the templates, using the \cite{Cardelli} law
(again with $E(B-V)$ in the range 0$-$0.5).


\subsection{SEDs for fitting spectra}
\label{sec_miles}

For each galaxy in the optical+NIR sample, we fit BC03/CB10 templates in the
wavelength range 3800$-$9200\AA~to its SDSS spectrum. Per other studies using
this sample, we also run {\it STARLIGHT} using SSPs from \cite[][hereafter, M09]{MILES} 
that are based on the MILES stellar library \citep{2006MNRAS.371..703S},
which has an almost complete coverage of stellar atmospheric parameters, containing
spectra of stars in the solar neighborhood. These SEDs cover the spectral range
3800$-$7500\AA~with a spectral resolution of 2.3\AA. Hence, they are well suited to analyzing SDSS
spectra, whose spectral resolution is $\sim$2.36\AA~(FWHM) in the wavelength
interval 4800$-$5350\AA. By improving on the instrumental homogeneity among stellar spectra, which share the same wavelength scale and resolution, and the accuracy of relative flux calibrations as compared to previous stellar population model catalogs, we expect M09 to contribute only small systematic biases in the estimates of spectroscopic parameters. Since support for the M09 library is not provided within {\it LePhare} by default, and given the minimal usable wavelength range, we have opted to use
BC03 models in the photometric SED fitting to compute the theoretical magnitudes, provided reliable
spectroscopic measurements are obtained from the M09 models.

To run {\it STARLIGHT}, we select a basis of 76 MILES SSPs, with ages and metallicities listed in
Table \ref{tab_ssp}. All models have [$\alpha/Fe$]$=0$ (i.e. solar abundance ratio). Dust extinction was applied to the templates, using the \cite{Cardelli} law.

\section{SED fitting using a $\chi^{2}$-minimization}
\label{sec_technique}

The theoretical SEDs are compared with either the observed galaxy colors or the 
observed spectra to determine the age, metallicity, and color excess $E(B-V)$ that 
best fits the observations.
In the case of galaxy colors, the stellar population models (see Section~\ref{sec_models}) can be used to produce theoretical SEDs at different redshifts, and hence the fitting
provides an estimate of the photometric redshift (see \cite{PhotoZReview} for 
a review).

\subsection{Fitting broadband photometry with {\it LePhare}}
\label{sec_41}

We perform the SED fitting with the photometric redshift code {\it LePhare} 
\citep{LePhare1, LePhare2}, which uses a $\chi^{2}$-minimization technique. Other popular
photometric redshift codes have been tested on various figures of merit.
\cite{2008arXiv0812.3831A} use $\sim$13,000 spectroscopic redshifts from
luminous red galaxies in SDSS DR6 to test six publicly available photometric
redshift codes, including {\it LePhare} and the popular codes \emph{HyperZ}
\citep{HyperZ} and \emph{BPZ} \citep{PhotoZReview}, finding that {\it LePhare}
performs best in the lower redshift intervals. This result and the existing
collaboration with the code developers motivates this choice for our study. Each
SED is redshifted up to $z=0.3$ in steps of $\delta z=0.005$ and convolved with
the SDSS/UKIDSS filter transmission curves \footnote{The SDSS filter curves were
obtained from
\url{http://www.sdss.org/dr6/instruments/imager/filters/index.html}, which
include instrument efficiency. UKIDSS filter curves are taken from
\cite{Hewett}. All transmission is considered at an airmass of 1.3.}. The
opacity of the intergalactic medium \citep{1995ApJ...441...18M} is taken into
account within {\it LePhare}. The merit function $\chi^{2}$ is defined as
\begin{equation}
\label{chi2}
	\chi^{2} (z,T,A) = \sum_{f=1}^{N_{f}} \left( \frac{F_{obs}^{f} - A
\times F_{pred}^{f} (z,T)}{\sigma_{obs}^{f}} \right) ^{2},
\end{equation} 
where $F_{pred}^{f}(T,z)$ is the flux predicted for a template $T$ at redshift
$z$. $F_{obs}^{f}$ is the observed flux and $\sigma_{obs}^{f}$ is the associated
error, converted from the AB system. The index $f$ refers to the specific filter
and $N_{f}$ is the number of filters. The photometric redshift is estimated from
the minimization of $\chi^{2}$ varying the three free parameters $z$, $T$, and
the normalization factor $A$. This normalization factor depends on the choice
of waveband used for scaling and is calculated as
\begin{equation}
\label{scaling}
	A = \sum_{\tilde{f}=1}^{N_{\tilde{f}}} \left( \frac{F_{obs}^{\tilde{f}}
\times F_{pred}^{\tilde{f}}}{ ( \sigma_{obs}^{\tilde{f}} )^{2}} \right) /
\sum_{\tilde{f}=1}^{N_{\tilde{f}}} \left(
\frac{F_{pred}^{\tilde{f}}}{\sigma_{obs}^{\tilde{f}}} \right)^{2},
\end{equation} 
where $\tilde{f}$ refers to the waveband(s) used for scaling. Studies often use
near-infrared bands (e.g. $K$-band) for scaling, since it is only weakly
affected by dust extinction and is quite insensitive to the presence of young,
luminous stars \citep{1984MNRAS.211..833L, 1995MNRAS.275L..19G, downsizing}. In
Section \ref{sec_photometry}, we present a comparison of galaxy stellar masses
for several choices of band scaling. Here, we note only that it produces a
negligible difference in the mass and, unless otherwise noted, we utilize all
available wavebands to scale the SED.

\subsubsection{Improving data and model matching with photometric zero-point
offsets}
\label{sec_zeropoint}

The SDSS/UKIDSS photometric zero-points are uncertain at the few percent level
(see \cite{Fukugita, Hewett}, respectively). To insure further accuracy of data
and model flux matching, it is important to provide reliable uncertainties in
the zero-point magnitudes. To this end, we utilize the complete and optical+NIR
samples with spectroscopic redshifts, including objects in the range $14.0 \le i
\le 16.4$ (brighter 25\% of \emph{ugrizYJHK} galaxies). Using a
$\chi^{2}$-minimization (Equation \ref{chi2}) at fixed redshift, we determine
for each galaxy the corresponding best-fitting stellar population template.
{\it LePhare} notes in each case $F_{obs}^{f}$, the observed flux in the filter $f$.
$A \times F_{pred}^{f}$ is the predicted flux derived from the best fit template
and rescaled using the normalization factor $A$ of Equation \ref{scaling}. For
each filter $f$, the sum
\begin{equation}
\label{zeropoint}
	\psi^{2} = \sum^{N_{gal}} \left( \left( A \times F_{pred}^{f} -
F_{obs}^{f} + s^{f} \right) / \sigma_{obs}^{f} \right)^{2}
\end{equation} 
is minimized, leaving $s^{f}$ as a free parameter. For random, normally
distributed uncertainties in the flux measurement, the average deviation $s^{f}$
should be zero. Instead, we observe some systematic differences, which are
listed for the optical+NIR sample at fixed redshift in Table \ref{tab_zeropoint} (median, converted to
magnitudes. In our data, these differences never exceed $0.042$ mag
(\emph{Y}-band) and have an average amplitude of $0.019$ mag, using BC03. We see
that these differences depend very weakly on the magnitude cut, listed as $14.0 \le m_{i}
\le x$, adopted to select
the sample and are also almost independent from the set of templates used in the fitting. The
sizes of these systematic differences are comparable to the level at which the
SDSS/UKIDSS photometric systems differ from a true AB system. We then proceed to correct
the predicted apparent magnitudes for
these systematic differences, using the correction factor $s^{f}$. If we repeat
a second time the procedure of template fitting after having adjusted the
zero-points, the best fit templates may change. {\it LePhare} checks that the
process is converging: after two iterations each estimated correction $s^{f}$
varies less than 0.0013 mag. Since the uncertainties in these zero-point
corrections are not less than 0.01 mag, this error is added in quadrature to
the apparent magnitude errors.

\begin{table}[htdp]
\caption{Systematic differences $s^{f}$ between the
observed and predicted fluxes.}
\begin{center}
\begin{tabular}{cccccc}

\hline
\hline
filter & BC03 & BC03 & BC03 & CB10 & PEGASE.2 \\
& $i \le$ 16.4 & $i \le$ 16.8 & $i \le$ 17.1 & $i \le$ 16.4 & $i \le$ 16.4 \\
\hline
$u$ & -0.021 & -0.026 & -0.026 & -0.020 & -0.161 \\
$g$ & 0.011 & 0.010 & 0.009 & 0.012 & 0.003 \\
$r$ & -0.006 & -0.004 & -0.004 & -0.004 & 0.050 \\
$i$ & -0.013 & -0.012 & -0.011 & -0.035 & -0.013 \\
$z$ & 0.004 & 0.005 & 0.005 & -0.017 & 0.019 \\
$Y$ & 0.042 & 0.042 & 0.041 & 0.062 & 0.093 \\
$J$ & -0.029 & -0.027 & -0.025 & -0.012 & -0.034 \\
$H$ & -0.021 & -0.023 & -0.025 & -0.025 & -0.049 \\
$K$ & 0.027 & 0.026 & 0.026 & 0.015 & -0.064 \\
\hline
\end{tabular}
\end{center}
\label{tab_zeropoint}
\end{table}

\subsection{Spectral fitting}

\begin{figure}[h]
\begin{center}

\includegraphics[width=6.5in]{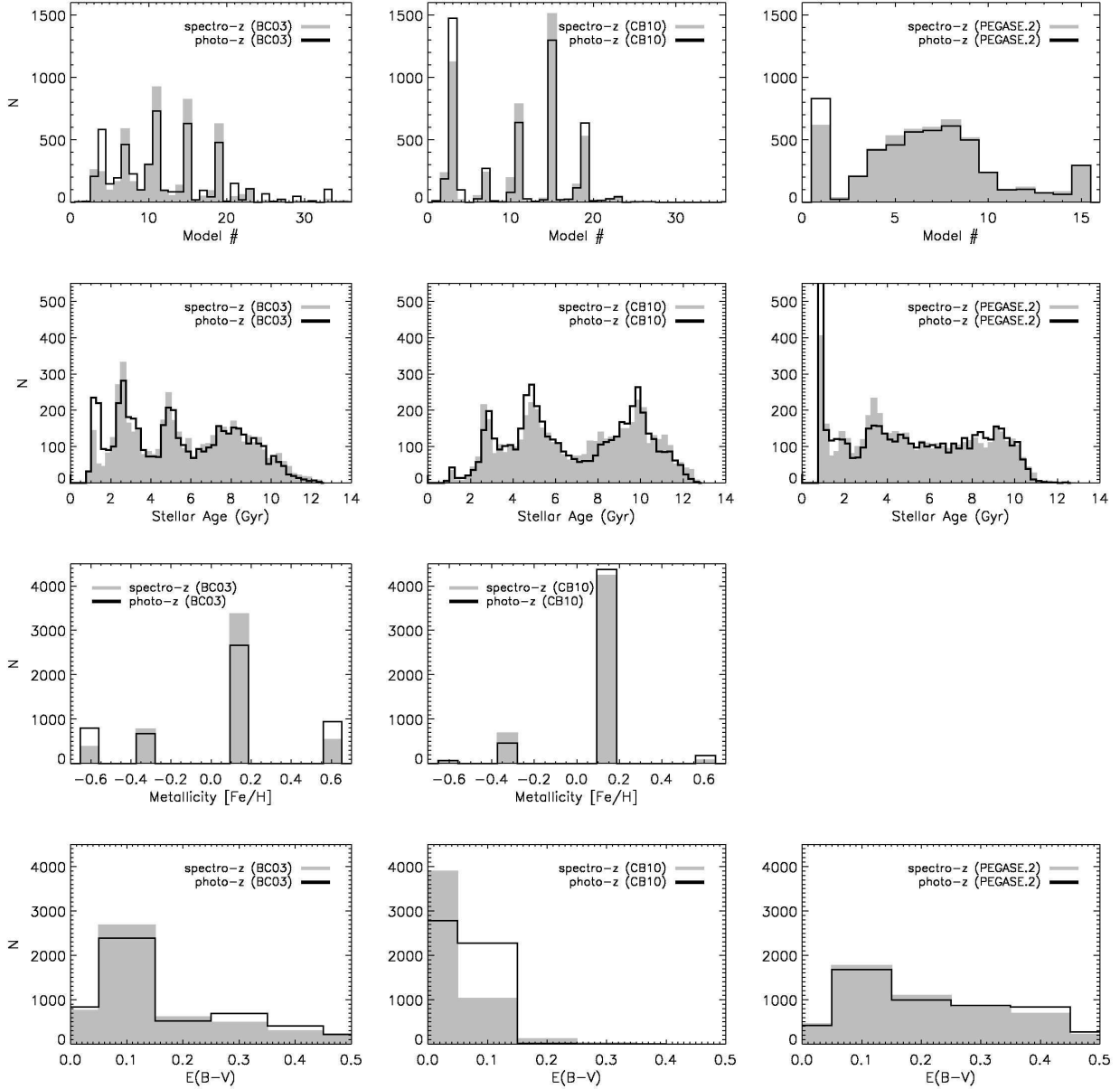}

\end{center}
\caption[XXX]
{\label{histograms_best_fit}
Distributions of models, ages, metallicities, and color excesses, using BC03
($left$), CB10 ($center$), and PEGASE.2 ($right$). See Table \ref{tab_models}
for ``Stellar Population Model \#'' information. $Gray$ histograms use the
spectro-z, and $black$ histograms use the photo-z. Note that PEGASE.2 templates
use solar metallicity. Stellar age is measured using the median of $\exp
(-\chi^{2}/2)$ from all fits, whereas {\it LePhare} only provides model
(metallicity) and $E(B-V)$ from the best $\chi^{2}$ value.
}
\end{figure}

For each spectrum in the optical+NIR sample, we run {\it STARLIGHT} \citep{STARLIGHT} using a suite of
models with stellar population parameters given in Table \ref{tab_ssp},
separately for BC03/CB10 and M09 (see Section~\ref{sec_miles}). The code uses 
a $\chi^{2}$-minimization, with a figure of merit given by
\begin{equation}
\label{chi2_starlight}
	\chi^{2} (\textbf{\emph{x}}, A_{V}, v_{*}, \sigma_{*}) = \sum_{\lambda} \left[
\left(\mathcal{O}_{\lambda} - \mathcal{T}_{\lambda} \right) \omega_{\lambda}
\right] ^{2},
\end{equation} 
where $\mathcal{O}_{\lambda}$ is the observed spectrum at wavelength $\lambda$,
$\mathcal{T}_{\lambda}$ is the model spectrum, and $\omega_{\lambda}$ is a
weighting factor. The model spectrum is a convolution of the intrinsic spectrum
$\textbf{\emph{x}}$ with a Gaussian filter centered at $v_{*}$ with dispersion $\sigma_{*}$.
Since both BC03/CB10 and M09 have similar spectral resolutions to the SDSS spectra,
the $\sigma_{*}$ provides a direct estimate of the galaxy velocity dispersion 
(see Paper I for details). The intrinsic spectrum $\textbf{\emph{x}}$ is a linear 
combination of a set of stellar population models (the {\it base}) provided as input 
to the code. The coefficients of this linear combination are computed from {\it STARLIGHT}
as part of the $\chi^{2}$-minimization procedure.
For our purposes, we do not use this feature of {\it STARLIGHT}, but rather
use the reduced-$\chi^{2}$ statistic provided by the code for each model 
in the {\it base} to compute the age, [Fe/H], and $A_{V}$ of the best fit 
(i.e. lowest $\chi^{2}$) stellar population model. For processing efficiency the code 
permits a maximum of 300 base models, so to allow a reasonably dense grid of
ages/metallicities (columns 3,4 of Table \ref{tab_ssp}), we chose to use single
($t=0$) burst models, which is justified by the mostly short-burst fits shown in
Figure \ref{histograms_best_fit} (see {\it top} panels). Here we plot the distribution of best fit
BC03/CB10/PEGASE.2 models for the optical+NIR sample, where the abscissa shows
increasing $\tau$ (i.e. decay time scale for star-formation), with every fourth
value being a constant metallicity -- for reference, the first five peaks
(BC03/CB10) represent solar metallicity models with $\tau \le 3$ Gyr. A
comparison of best fit ages, metallicities, and extinctions between {\it LePhare}
and {\it STARLIGHT} is given in Section \ref{sec_svl}. 

\section{SED fitting reliability}
\label{sec_reliability}
In this section, we compare results obtained by fitting broadband colors and 
spectra of ETGs.
We begin by comparing photometric and spectroscopic redshifts in
Section~\ref{sec_photoz_comp}, while in Section~\ref{sec_svl} we compare age,
metallicity, and internal extinction estimates between {\it LePhare} and
{\it STARLIGHT}.

\subsection{Accuracy of photometric redshifts}
\label{sec_photoz_comp}
\begin{figure}[h]
\begin{center}

\includegraphics[width=6.5in]{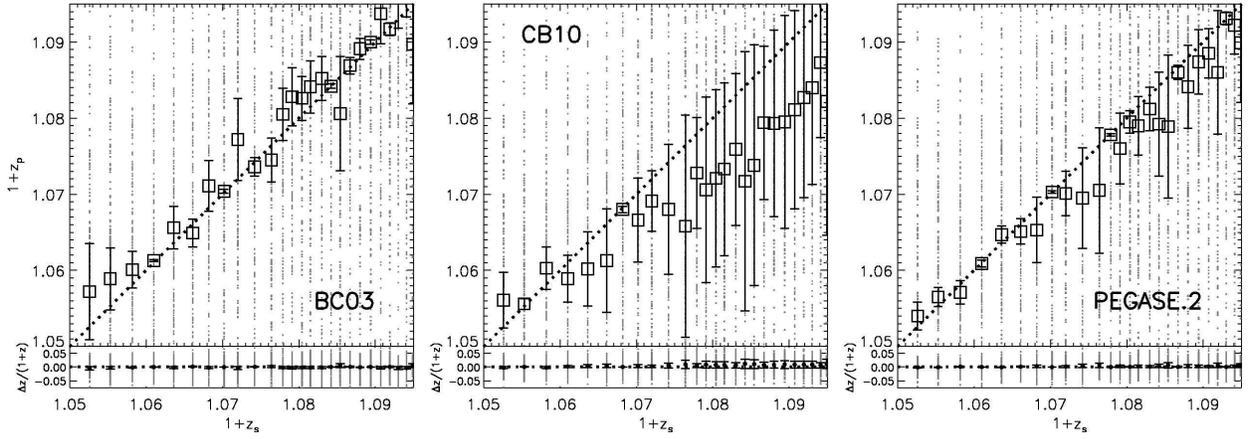}

\end{center}
\caption[XXX]
{\label{fig_redshifts}
Comparison between the spectroscopic and photometric redshifts, having corrected
for zero-point magnitude offsets (see Section \ref{sec_zeropoint}) with a bright
subsample $14.0 \leq m_{i} \leq 16.4$. $Top$ panels show redshifts for the optical+NIR sample, with $squares$ representing the median values in each bin (200 objects) and their 2$\sigma$ uncertainties. $Bottom$ panels show residuals, $z_{s}-z_{p}$, normalized by the spectro-z.
}
\end{figure}

We compare spectroscopic and photometric redshifts for the optical+NIR sample in
Figure \ref{fig_redshifts}. The galaxies are binned to include 200 objects per
$\delta z_{spec}$ bin, and the error bars denote the dispersion defined by $1.48
\times \mathrm{median}(|\Delta z|/(1+z_{spec}))$. This measurement of the
scatter corresponds to the rms for a Gaussian distribution and is unaffected by
outliers \citep{LePhare2}. We recover $\sim$73\% of the galaxies
(BC03) with outlier rate, defined here as $\eta = |\Delta z| <
0.025(1+z_{spec})$. The accuracy of these photometric redshifts
improved with addition of the \emph{u}-band data by $\sim$7\% in the outlier
rate. \cite{LePhare2} use their sample of $\sim$3,000 galaxies to confirm the
importance of the \emph{u}-band for low redshift galaxies, recovering $\sim$80\%
of the photometric redshifts at $z<0.4$ and $\sim$95\% using the \emph{u}-band.
Figure \ref{fig_redshifts} shows the binned redshifts, using the Chabrier
(Scalo) IMF for BC03/CB10 (PEGASE.2) stellar population models, and the Cardelli
extinction law. Photometric redshifts are well recovered, given the small
redshift range, with outlier rates of $\sim$73\%, $\sim$75\%, $\sim$76\% for
BC03/CB10/PEGASE.2, respectively. However, at higher redshifts, CB10 models
produce systematically lower photometric redshifts. This is
surprising given the improvements described in \cite{TP-AGB}, although it does not change the results that follow, which use the spectroscopic
redshift. We note that the extremely small redshift range of our sample makes these photo-z tests unsuitable for drawing conclusions for typical imaging surveys.

\subsection{Comparing photometric \& spectroscopic measurements}
\label{sec_svl}

To compare photometrically and spectroscopically determined
results, we use the same model parameters between {\it LePhare} and
{\it STARLIGHT} as much as possible. 
All theoretical parameters are consistent
between the two packages, except for the interstellar extinction, which is
calculated on-the-fly within {\it STARLIGHT}.
The comparison assumes that there is essentially no stellar 
population gradient between the spectroscopic (fiber) and 
photometric (Kron) apertures. This assumption is motivated by the lack of 
color gradients between fiber and Kron apertures, as shown in 
Appendix~\ref{app:fiber_kron_mags}.
As shown in Figure~\ref{fig_starlight_v_lephare_all}, 
there is good agreement between the distributions of photometric and spectroscopic 
ages using CB10 (see {\it bottom-left} panel), with a median difference of 0.75 Gyr, whereas the sample median
age (measured from AGE\_MED in {\it LePhare}) \footnote{Stellar ages and masses are
measured from the median of the likelihood function $\exp (-\chi^{2}/2)$ rather
than the best $\chi^{2}$. This reduces stochastic mass errors due to individual
models which happen to have very low $\chi^{2}$ values, since there remains some
degeneracy in the SED fits even with 9 filters. We also find that this method
produces masses that are more consistent with other studies.} estimated from the
BC03 templates is 1 Gyr too low, slightly improved by addition of the
\emph{u}-band data. However, the new treatment of the TP-AGB phase yields
isochrones up to 1 magnitude brighter in the \emph{K}-band \citep{TP-AGB}, where
the dominant flux of these galaxies is expected -- this would produce better
fits between the photometric SED and older stellar populations. Despite this
discrepancy, $ugrizYJHK$ photometry reproduces the age-metallicity distribution
much more accurately than the $ugriz$-bands alone, as expected, and so is used
in the tests that follow. Both {\it STARLIGHT} and {\it LePhare} produce similar
age/metallicity/extinction distributions for our optical+NIR sample, despite
large differences for individual galaxies and strong systematic trends as a function
of {\it STARLIGHT} output parameters (see Figure
\ref{fig_starlight_v_lephare_all_delta}).

\section{Method to estimate galaxy stellar masses}
\label{sec_method}

In most cases, all 9 bands were used to compute the stellar mass. We provide the
rescaled template stellar mass (measured from MASS\_MED in {\it LePhare}) derived
from
\begin{equation}
\label{eqn_m}
	\log \mathcal{M} = \log (\mathcal{M}_{*}/\mathcal{L}_{\lambda}) + 0.4
kcor_{\lambda} + 2 \log d_{pc} - 2.0
	+ 0.4 M_{\lambda}^{sun} - 0.4 m_{\lambda^{\prime}},
\end{equation} 
where $\mathcal{M}_{*}/\mathcal{L}$ is the stellar mass-to-light ratio in the
chosen filter centered at $\lambda$, $kcor_{\lambda}$ is the k-correction,
$d_{pc}$ is the cosmology dependent luminosity distance in pc, and
$m_{\lambda^{\prime}}$ is the apparent magnitude in the observed filter centered
at $\lambda^{\prime}$ (see \cite{2009MNRAS.394..774L} for a derivation of
Equation \ref{eqn_m}). As with stellar age, this measurement is taken from the
median of the likelihood function $\exp (-\chi^{2}/2)$. Note that stellar masses
computed in this manner are only partial due to the adaptive aperture of $3
\times r_{K,i}$; total stellar masses would require an aperture correction, which 
is irrelevant for this type of study (however, see Section \ref{sec_mpi}). 
We then compute the median and its
asymmetric 2$\sigma$ uncertainties for histograms of $\Delta ( \log \mathcal{M}
)$ for differences resulting from the choices of stellar population model, initial mass function,
interstellar extinction law, and photometric bands.

We also determine stellar mass, using the optical+NIR sample, for
fixed age, metallicity, and extinction, derived from {\it STARLIGHT}, in Sections
\ref{sec_ext}-\ref{sec_agemet}. We define the difference $\Delta$ as $\log
\mathcal{M}_{fixed} - \log \mathcal{M}_{free}$ for each parameter.
Operationally, $fixed$ refers to constraining the specified parameter and
leaving all other parameters (except for redshift) free in the
$\chi^{2}$-fitting. This allows us to quantify the accuracy of a stellar
mass estimate when these parameters are unknown, and then derive corrections for
any offset. Therefore, we define a \emph{mass correction} term, $\epsilon$,
which can be applied to the stellar mass measurements over a range of
photometrically determined ages and metallicities, for a given best fit template
(i.e. through a linear regression line). This term is computed as the median of
each $\Delta ( \log \mathcal{M} )$ histogram described above for fixed-parameter
minus free-parameter. In Section \ref{sec_age}, we discuss the use of this
parameter in correcting the stellar mass distribution of our samples.

Lastly, we present two comparisons of our measured stellar masses to similar estimates from a group at the Max Planck Institute and to those presented in Paper IV -- this is discussed in Section \ref{sec_mpi}. We note here that the overall agreement with separate photometrically estimated stellar masses (MPI) is excellent, showing a negligible offset with 2$\sigma$ scatter of about 0.1 dex. We find that inclusion of short-burst model SEDs when fit with a young ($\lesssim$3 Gyr) stellar population yield stellar masses that are in most cases $\sim$0.2 dex too low.

\section{Biases and uncertainties in stellar masses}
\label{sec_results}

Unless otherwise noted, the results presented herein use the optical+NIR sample
of 5,068 ETGs and their spectroscopic redshifts.  In Sections \ref{sec_mc} and
\ref{sec_photoz}, we test the impact of sample photometric errors and redshifts,
respectively, on the resulting stellar mass estimates. The tests presented in
Sections \ref{sec_photoz}-\ref{sec_ext} were conducted separately for both BC03
and CB10 stellar population models. In Sections \ref{sec_ext}-\ref{sec_agemet},
we apply results from the {\it STARLIGHT} spectral fits to constrain the
$\chi^{2}$-minimization, using the BC03, CB10, and M09 libraries, separately for
each test. When spectroscopically constraining $t$, $Z$, or $A_{V}$ with the M09
library, we use BC03 models within {\it LePhare}, as described in Section
\ref{sec_miles}. In Table \ref{tab_summary}, we also summarize the results of
the following tests for BC03/CB10 stellar population models. Similar results
using the complete sample of ETGs are provided in Table \ref{tab_summary_complete}. The reader should take results from this sample with caution, considering the caveats introduced by use of only five optical bandpasses and the discussions in Section \ref{sec_svl} and below. Further, we do not perform the SED fitting for fixed age/metallicity/extinction on the complete ($ugriz$-only) sample due to resulting catastrophic errors in the constrained $\chi^{2}$-minimization.

\subsection{Dependence on sample magnitude errors}
\label{sec_mc}

\begin{figure}[h]
\begin{center}

\includegraphics[width=4in]{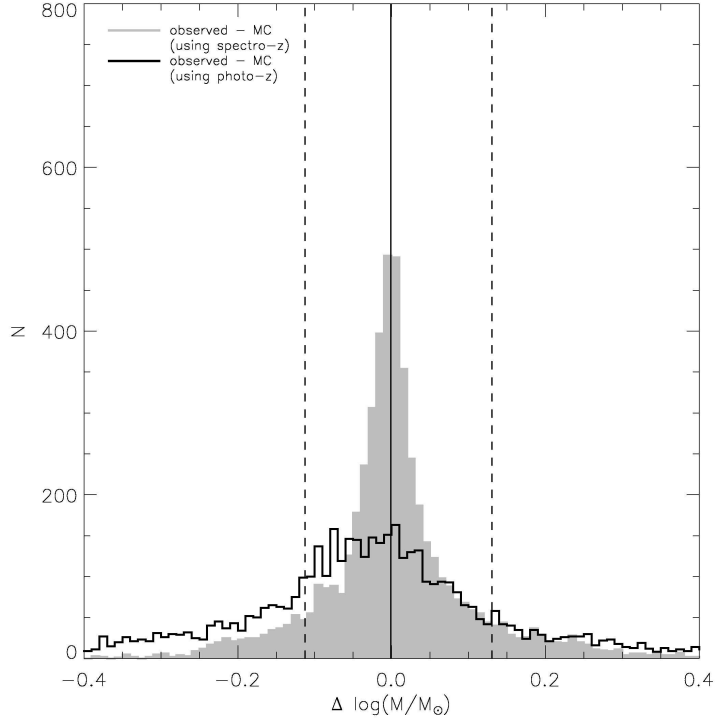}

\end{center}
\caption[XXX]
{\label{fig_MC}
Errors in the stellar mass computed from the observed apparent magnitudes minus
the Monte-Carlo magnitudes (see Section \ref{sec_mc}), using the
spectro-z (\emph{gray}) and photo-z (\emph{black}), for the optical+NIR sample. The $ugrizYJHK$ magnitudes
were varied within their 1$\sigma$ errors according to a normal distribution,
and the resulting magnitudes were used to scale the template SED to obtain the
stellar mass estimate. The \emph{solid} and \emph{dashed} lines indicate
median and $\pm2\sigma$ limits, respectively, for the $gray$ histogram.
}
\end{figure}

The photometric measurements of the complete and optical+NIR samples in this study
have high S/N ratios (see Figure \ref{fig_SNR}). To investigate the
dependence of galaxy stellar mass estimates on photometric errors, we used a
Monte-Carlo approach to varying the $ugrizYJHK$ magnitudes. Each magnitude was
varied according to a normal distribution, centered on the observed apparent
magnitude, using the $1\sigma$ error limits provided by SExtractor. For fixed
redshift, the resulting differences in stellar masses from the original sample (optical+NIR) are
shown in Figure \ref{fig_MC}. The masses show a dispersion of
$\pm 0.12$~dex at the 95\% CL, which lie within uncertainties induced
by the photometric redshift, but are surprisingly large given the S/N of this
sample, with a negligible systematic bias introduced in the mass.

\subsection{Dependence on photometric redshift}
\label{sec_photoz}

\begin{figure}[h]
\begin{center}

\includegraphics[width=6.5in]{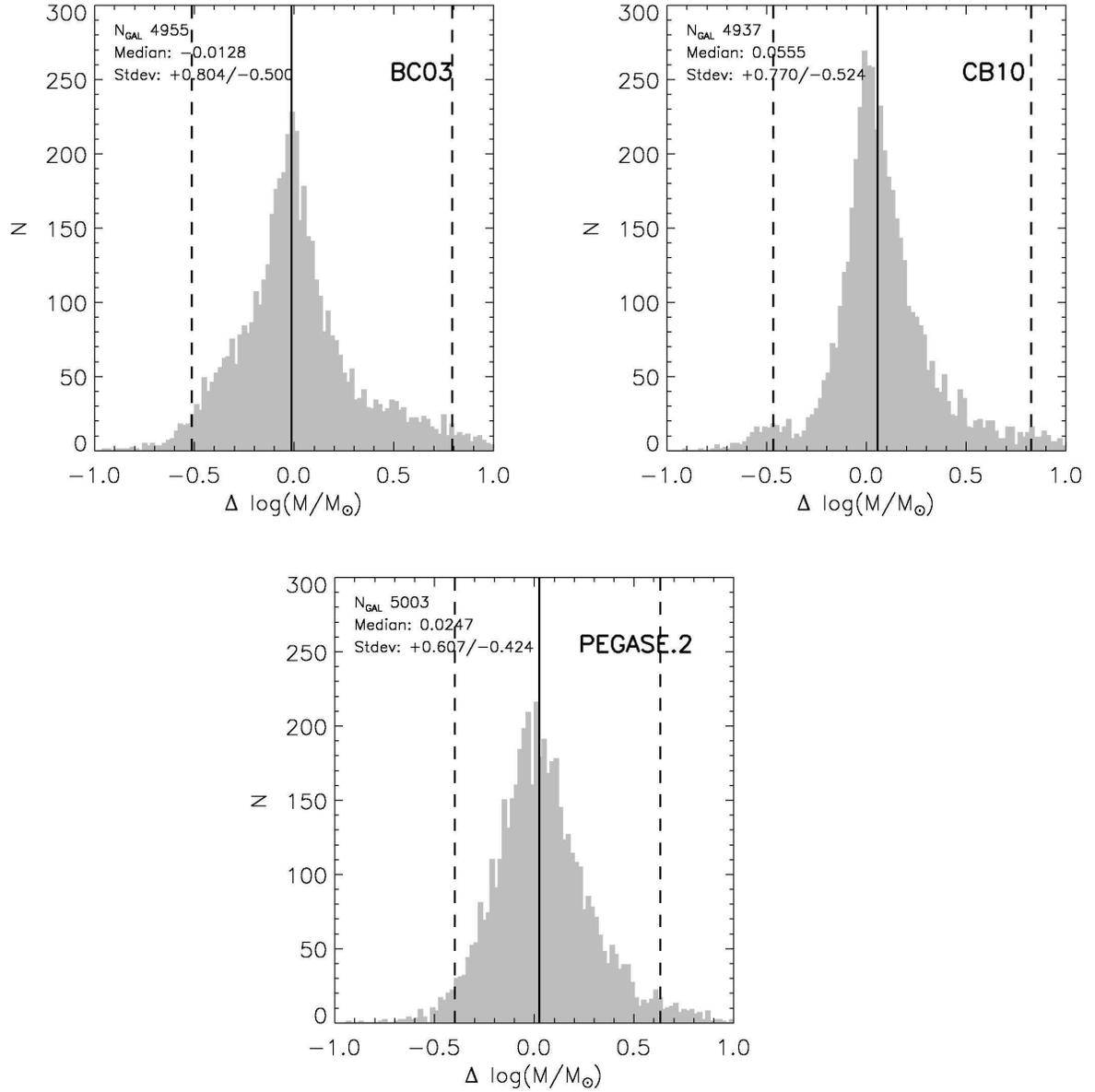}

\end{center}
\caption[XXX]
{\label{fig_mass_redshift}
Histograms of $\Delta ( \log \mathcal{M} )_{s-p}$, defined in Section
\ref{sec_method}, for spectroscopic and photometric redshifts, using the BC03,
CB10, and PEGASE.2 stellar population models. The \emph{solid} and
\emph{dashed} lines indicate median and $\pm2\sigma$ limits, respectively. Only objects
in the range $9 < \log \mathcal{M}/\mathcal{M}_{\sun} < 12$ are included in this figure.
}
\end{figure}

We quantified how the stellar mass accuracy is affected by the use of a photometric redshift, which is the typical case for a large imaging survey.
Figure \ref{fig_mass_redshift} shows the difference between the stellar masses
computed with the spectroscopic vs. photometric redshifts. The sample uses the entire redshift
range $0.05 \leq z \leq 0.095$, and all bands are used
to scale the mass. We find a median difference of $\sim$0.01 dex, with errors
of $+0.804$/$-0.500$ dex at the 95\% CL, using BC03 models. This scatter is
slightly larger than the systematic uncertainties expected (0.2 dex at the 68\%
CL) in the stellar mass estimate due to photometric redshifts
\citep{2007A&A...474..443P, 2009MNRAS.394..774L} at low redshift
\footnote{\cite{2007A&A...474..443P} find uncertainties of $\sim$0.2 dex at $z <
0.4$, with increasing errors at lower redshifts. Our redshift range is much
lower, so these errors may be expected.}. For the optical+NIR sample, the
average positive 2$\sigma$ limit on the photo-z is roughly 40\% -- at the
largest redshift ($z=0.095$, i.e. $d_{L}=435$ Mpc), this yields a luminosity
distance error of nearly 200 Mpc, or a stellar mass error (from Equation
\ref{eqn_m}) of $\sim$0.4 dex. The photo-z also impacts the template best fit
ages differently, depending on stellar population model. We concluded that the unusually low redshift of our sample results in abnormally large mass errors (compared to those reported in the literature) due to the use of photo-zs, since the fractional errors in the luminosity distances are large. For all remaining tests, we use the spectroscopic redshifts.

\subsection{Dependence on stellar population model}
\label{sec_ssp}

The comparison of stellar masses between different models is shown in Figures
\ref{fig_histogram_masses_BC03} and \ref{fig_histogram_masses_CB10} (see {\it top-left} panels).
To allow a meaningful comparison with BC03/CB10 and PEGASE.2 models, 
here we include only objects in the optical+NIR sample that are fit with a solar
metallicity ($N_{BC03}=3,371$, $N_{CB10}=4,240$) \footnote{These results assume
a \cite{Scalo86} IMF for PEGASE.2 models. To compare with BC03/CB10 results,
which use a Chabrier IMF, we remove the bias in stellar mass introduced by the
IMF, as described in the Appendix~\ref{app:imf}}.
The stellar mass comparisons between models yield small offsets. CB10 models
produce slightly higher masses by about 0.005 dex, with a scatter of
$\sim$0.14 dex with respect to BC03 masses using the spectroscopic redshift 
(see Table~\ref{tab_summary}). This result seems to be in disagreement with, instead, BC03 models
having systematically higher NIR $\mathcal{M}_{*}/\mathcal{L}$ than CB10 models, at given
age and metallicity (see e.g. \citealp{2008MNRAS.384..930E}). Indeed, $LePhare$ gives lower photometric ages when using BC03 (see Figure \ref{fig_starlight_v_lephare_all}), resulting in
lower $\mathcal{M}_{*}/\mathcal{L}$ ratios when compared to CB10 results. This might compensate 
for any variation of $\mathcal{M}_{*}/\mathcal{L}$ due to the different treatment of the TP-AGB phase between BC03 and CB10. This seems to hold only for ETGs with old ages, as \cite{2009ApJ...701.1839M} actually found CB10 (as of 2008) based stellar masses to be systematically smaller than those obtained
from BC03 for high-redshift ($z \sim 2.3$) galaxies. Moreover, \cite{2006ApJ...652...97V} reported
that, for $z \sim 0$ galaxies, including the TP-AGB phase, as done in \cite{Maraston05} stellar
population models, leads to lower stellar masses than those obtained with BC03, when fitting
both optical and NIR broad-band colors, more consistent with what is reported here. On the other
hand, PEGASE.2 models produce systematically lower masses than BC03/CB10. The offsets in
these comparisons are likely due to the different prescriptions for the luminous stars dominating
the $K$-band flux, and are comparable to other factors discussed below.
\cite{2009MNRAS.394..774L} measure the effect of different TP-AGB prescriptions (assuming
these stars dominate the $K$-band flux) between stellar population models on the resulting stellar
mass when the age is known, finding agreement within at least 20\%, between BC03,
CB10, and PEGASE.2 models.

Earlier studies \citep{2007A&A...474..443P, IMF_ETGs} find that a Chabrier IMF
underestimates galaxy stellar masses by nearly a factor of 2, compared to those
derived with the Salpeter IMF. \cite{2006MNRAS.366.1126C} used different models
than these authors and find that galaxy stellar masses based on a Salpeter IMF
were in some cases too high compared to those determined with stellar
kinematics, reaching the same conclusion.
We use the stellar population
models of BC03/CB10 to further compare model dependent stellar masses (with
star-formation histories, metallicities, and color excesses given in Table
\ref{tab_ssp}), using different IMFs. Figures \ref{fig_histogram_masses_BC03} and
\ref{fig_histogram_masses_CB10} show the dispersion in $\Delta (\log
\mathcal{M})$ between the Chabrier and Salpeter IMFs (see {\it top-right} 
panels). The zero-point offset is $-0.227$ ($-0.225$) dex using BC03 (CB10) models, or a factor
of $\sim$2, which differs from the analytical estimate (see Appendix~\ref{app:imf}) by $\sim$0.1 dex.


\subsection{Dependence on photometric bandpasses fitted}
\label{sec_photometry}

Here we highlight the systematic effects of band scaling in the
$\chi^{2}$-minimization as well as inclusion of ultraviolet and NIR continua,
separately, in the SED fitting. Again, we remind the reader that these results
apply specifically to a low redshift population of ETGs, where the continuum
levels behave differently with redshift for ETGs and especially for other
morphological types.

\begin{figure}[h]
\begin{center}

\includegraphics[width=4in]{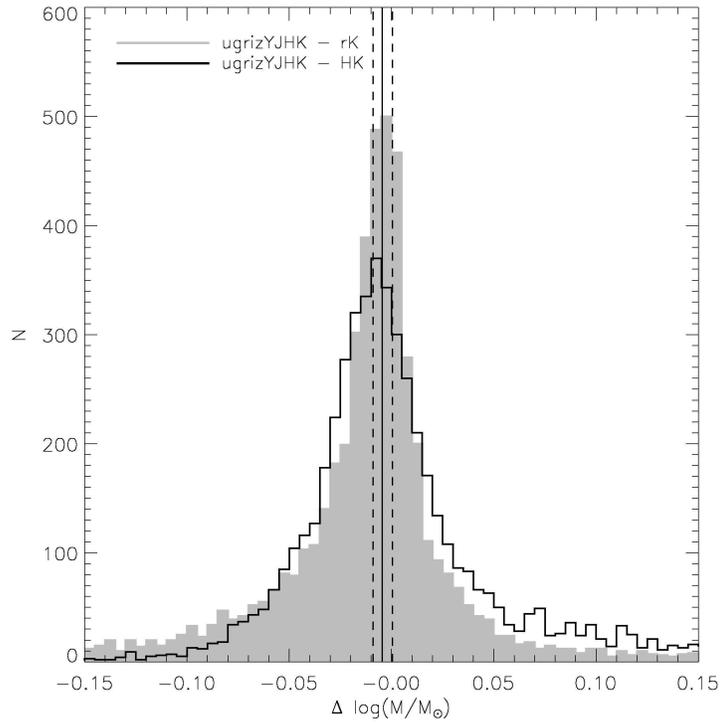}

\end{center}
\caption[XXX]
{\label{fig_scaling}
Effect of using different optical+NIR bandpasses for scaling the SED (using
BC03). With known redshift, this histogram suggests that there is no systematic
bias in choosing all bands as opposed to $K$-band to scale the SED. The median
difference (\emph{solid vertical}) is indistinguishable from zero at the $0.3\sigma$
level (\emph{dashed vertical}). Note that {\it LePhare} does not accept only 1-band for
scaling when the redshift is unknown.
}
\end{figure}

In Figure \ref{fig_scaling}, we plot the differences in galaxy stellar mass for
several choices of band scaling -- $ugrizYJHK$, $rK$, and $HK$. For our
optical+NIR sample of low redshift ETGs, this allows for a measurement of the
sensitivity of the overall luminosity scaling of the SED to magnitude errors,
specifically targeting the $K$-band. The median, dispersion, kurtosis, and
skewness for these stellar mass differences are given in Table
\ref{tab_scaling}, suggesting that there is a systematic bias in the stellar
mass of 0.005 dex when using all bands as opposed to using only the $H$ and $K$ bands. The
differences are sharply peaked around zero, as indicated by the high excess
kurtosis.

\begin{table}[htdp]
\caption{Characteristics of $\Delta ( \log \mathcal{M} )$ for $ugrizYJHK-xK$ (with $x=r, H$) band scaling.}
\begin{center}
\begin{tabular}{ccccc}

\hline
\hline
& Median$^{+2\sigma}_{-2\sigma}$ & Kurtosis & Skewness \\
\hline
$ugrizYJHK-rK$ & $-8.000$E-$3^{0.06041}_{0.09934}$ & 9.09 & -0.734 \\
$ugrizYJHK-HK$ & $-4.500$E-$3^{0.09971}_{0.04286}$ & 5.61 & 1.00 \\
\hline

\end{tabular}
\end{center}
\label{tab_scaling}
\end{table}

These offsets are small compared to errors introduced by other factors explored
later in this paper. The excess of positive residuals for $ugrizYJHK-HK$ scaling
is possibly due to a known poor modeling effect in the near-infrared described in
Sections \ref{sec_reliability} and \ref{sec_ssp}, which
predicts $K$-band magnitudes that are too faint. Therefore, we use all bands to
measure the stellar mass, so that the scale factor is less sensitive to errors
in any one given band. In the minimization, the best $\chi^{2}$ at each redshift
step is saved to build the function $F(z) = \exp [-\chi_{min}^{2}(z)/2 ]$. This
function is used to refine the photo-z solution with a parabolic interpolation
\citep{1969drea.book.....B}.

\subsection{Dependence on inclusion of rest-frame UV photometry}
\label{sec_u}

We use the optical+NIR sample to compare stellar masses computed with and
without \emph{u}-band data, for BC03/CB10 models. We find scatters of $\sim$0.14
and $\sim$0.12 dex, respectively, in $\Delta ( \log \mathcal{M} )$, as shown in
Figures \ref{fig_histogram_masses_BC03} and \ref{fig_histogram_masses_CB10},
with negligible systematic bias in the mass (see {\it green} histograms in
{\it bottom-right} panels). \cite{2010ApJ...712..833C} describe
the influence of the 4000\AA~break (quantified as D$_{n}$4000) on $ugr$ colors
for red sequence galaxies and find that BC03 models, among others, predict
D$_{n}$4000 strengths that are generally too large. It is possible that this
effect, which alone would yield older stellar ages in the $\chi^{2}$-fitting,
competes with the underpredicted NIR luminosities to produce the scatter
observed in Figure \ref{fig_histogram_masses_BC03} (for $u-$no-$u$). This
uncertainty would be difficult to quantify, as these models suffer in
reproducing the age-metallicity plane found from spectral fits. However, with
the modifications in CB10 models affecting spectra redward of $\sim$6000\AA,
the $u-$no-$u$ stellar mass differences in Figure
\ref{fig_histogram_masses_CB10} have a lower systematic bias and dispersion than
the BC03 models, showing that CB10 treats optical+NIR spectral regions more
consistently with the observations. Hence, including $u$-band data simply provides
more robust stellar mass estimates.

\subsection{Dependence on rest-frame NIR photometry}
\label{sec_YJHK}

We also test the absence of near-infrared data on the stellar mass estimate (see {\it bottom-right} panels of Figures \ref{fig_histogram_masses_BC03} and 
\ref{fig_histogram_masses_CB10}). The
$\Delta ( \log \mathcal{M} )$ offset in $ugrizYJHK-ugriz$ stellar masses is
approximately $-0.057 \pm 0.10$ ($-0.042 \pm 0.12$) dex for BC03 (CB10) models,
with $ugriz$-only photometry producing higher stellar masses in both cases.
\cite{2007A&A...474..443P} find a similar result -- that is, higher masses by
$\sim$0.1 dex using optical-only photometry -- for their $K$-selected
spectroscopic sample of high redshift galaxies. Given the changes in TP-AGB
evolution between BC03/CB10 models, these relatively smaller dispersions suggest
that the 4000\AA~break is a more sensitive constraint in the SED fits than is
the continuum level of the NIR data for this sample of low redshift ETGs.

\subsection{Dependence on interstellar extinction}
\label{sec_ext}

We compare stellar masses derived using the \cite{Cardelli} and \cite{Calzetti}
extinction laws with $E(B-V)$ values listed in Table \ref{tab_ssp}. These
empirical laws are line-of-sight dependent with average $R_{V}$ equal to $3.1$ and $4.05
\pm 0.80$, respectively, in popular photometric redshift codes (e.g.
$BPZ$, $HyperZ$, {\it LePhare}). Using BC03 models, this mass difference yields
errors of $+0.07$/$-0.14$ dex for the optical+NIR sample with negligible
systematic bias in the mass. These extinction laws produce excellent mass
agreement, however, using CB10 models, which might be related to the fact that CB10 models provide 
SEDs that are more consistent with the observations in both the optical and NIR.
We also measure stellar masses using the \cite{CF2000} extinction law, which is provided
with the $GALAXEV$ code and requires as inputs the total effective $V$-band
optical depth $\tau_{V}$ and the fraction $\mu$ of it contributed by the
interstellar medium. The default values of $\tau_{V}=1.0$ and $\mu=0.3$ are
assumed for this test, where it should be noted that these values best
describe a population of star-forming galaxies. Our optical+NIR sample yields a
$\Delta ( \log \mathcal{M})$ zero-point offset of $-0.10$ dex for BC03 models,
with CB10 models producing slightly better agreement. Note that this difference is comparable to the offsets found by \cite{COSMOS} for the $z \sim 1$ zCOSMOS sample
($\Delta ( \log \mathcal{M})_{\mathrm{(Calzetti}-\mathrm{CF2000)}}$ is $-0.14$ dex), where they
also show that this systematic offset is larger for massive galaxies with a high SFR.

\begin{figure}[h]
\begin{center}

\includegraphics[width=6.5in]{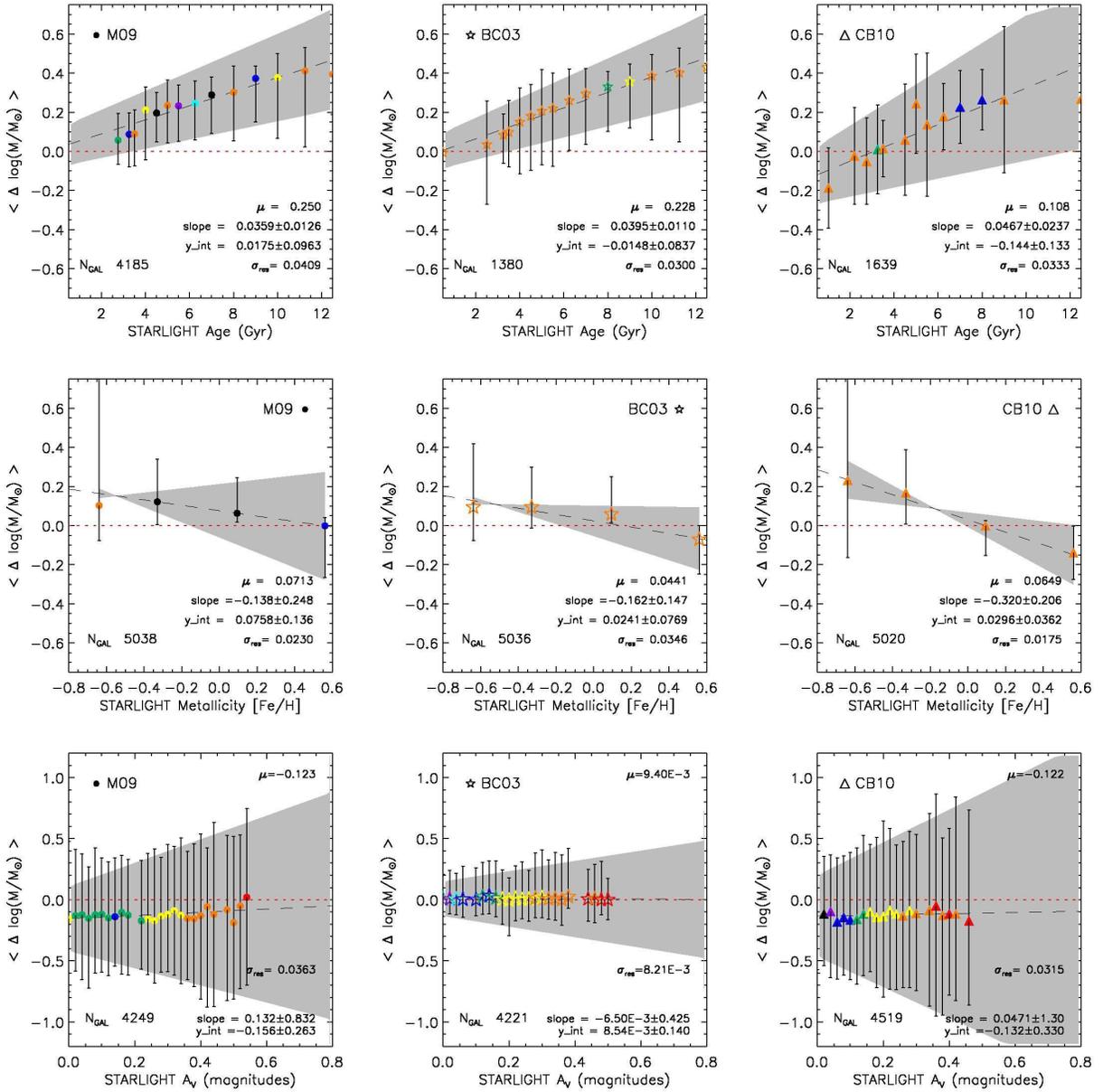}

\end{center}
\caption[XXX]
{\label{fixed_histograms_opticalNIR}
Plots showing $\log \mathcal{M}_{fixed} - \log \mathcal{M}_{free}$ offsets, described in Section \ref{sec_method}, as a function of spectroscopically determined age, metallicity, and extinction for the optical+NIR sample, using M09 ($left$), BC03 ($center$), and CB10 ($right$) models. The color code is -- from \emph{red} to \emph{blue} -- \# galaxies $<50$, $<100$, $<200$, $<300$, $<400$, $<500$, $<600$ (\emph{black}~$>600$). Measurements in a given bin are only considered if the bin contains $\geq$40 galaxies without catastrophic fitting failures (as returned by $LePhare$). Overplotted are linear fits to each data set. The error bars denote the 95\% CL on the median, normalized by the square root of the number of galaxies in the given bin. Provided within each plot are the mean offset ($\mu$), the standard deviation of the residuals ($\sigma_{\mathrm{res}}$), and the slope and y-intercept for a model line fit from a $\chi^{2}$-minimization.
}
\end{figure}

Finally, we utilize spectroscopic measurements of $A_{V}$ (from {\it STARLIGHT}) and the
\cite{Cardelli} extinction law to constrain the interstellar extinction for each
galaxy. In {\it LePhare}, we fix the color excess $E(B-V)$ based on the relationship
$E(B-V)=A_{V}/R_{V}$. We find that for our optical+NIR sample, the uncertainties in the
stellar mass estimate due to unknown extinction are comparable ($\sim$0.2 dex
for BC03) to the effect of using different extinction laws. However, M09/CB10 models yield
much larger uncertainties of $\sim$0.4 dex. We find a
negligible systematic trend in mass difference as a function of
spectroscopically determined extinction for all models. Figure \ref{fixed_histograms_opticalNIR} shows however that when fixing the extinction
to the spectroscopic value measured from M09/CB10 models, stellar masses are
systematically lower than the free parameter case by $\sim$0.15 dex for all
$A_{V}$ values. A linear regression fit to the trend of $\Delta (\log
\mathcal{M})$ offset with amount of extinction (in magnitudes) produces

\begin{equation}
\Delta (\log \mathcal{M}) = \left\{
	\begin{array}{ll}
		(0.132 \pm 0.832) A_{V} + (-0.156 \pm 0.263) & \mathrm{for}~ \mathrm{M09}, \\
		(-6.50\mathrm{E-}3 \pm 0.425) A_{V} + (8.54\mathrm{E-}3 \pm 0.140) & \mathrm{for}~ \mathrm{BC03}, \\
		(0.0471 \pm 1.30) A_{V} + (-0.132 \pm 0.330) & \mathrm{for}~ \mathrm{CB10} \\
	\end{array}
\right.\end{equation}
where the 1$\sigma$ uncertainties on the regression coefficients are provided above. Large uncertainties in $\Delta ( \log \mathcal{M} ) $ affect the reliability of this fit, but it is provided for reference.

Fitting broadband photometry seems to produce lower interstellar extinction than
spectral fitting for our sample of ETGs, as shown in the {\it right} panels in Figure
\ref{fig_starlight_v_lephare_all_delta}. This would lower the $0.4 \times
m_{\lambda^{\prime}}$ term in Equation \ref{eqn_m} and produce a higher stellar
mass estimate. We note here that we do not provide a \emph{mass correction}
analysis as a function of photometrically determined extinction, due to the
shallow trend in $\Delta (\log \mathcal{M})$ with $A_{V}$. Furthermore, the
$A_{V}$ (or equivalently $E(B-V)$) parameter space cannot be specified directly
within the {\it STARLIGHT} $\chi^{2}$-fitting.

\subsection{Dependence on stellar age and metallicity}
\label{sec_agemet}


This section spotlights both the effect of different best fit ages and
metallicities on the resulting stellar masses, and suggests how blue continuum
fluxes and mass-to-light ratios could explain the large stellar mass differences
due to photometrically determined ages. To this end, we have used optical spectra
to constrain independently age and metallicity in the models.

\subsubsection{Dependence on stellar metallicity}
\label{sec_met}

The Padova 1994 stellar evolutionary tracks used by BC03/CB10 encompass six
metallicities in the range $Z/Z_{\sun}=0.005-2.5$, with the four most metal-rich
models chosen for this study. The M09 model fits use $Z/Z_{\sun}=0.2, 0.4, 1.0,
1.6$ metallicities from the Padova 2000 library. We separate galaxies on the
basis of the metallicity ranges given in Table \ref{tab_ssp} and run {\it LePhare}
with the metallicity constrained to the closest value. The result is a systematic offset for unknown
metallicity consistent with zero for most objects fit with M09 models. However,
metallicity is more poorly constrained with BC03/CB10 models, yielding a larger
systematic offset in the stellar mass, with uncertainties as high as $\sim$0.2
dex at low metallicity (see Table~\ref{tab_summary}). The linear regression fit to our 
optical+NIR sample yields the following functional form:
\begin{equation}
\Delta (\log \mathcal{M}) = \left\{
	\begin{array}{ll}
		(-0.138 \pm 0.248) Z + (0.0758 \pm 0.136) & \mathrm{for}~ \mathrm{M09}, \\
		(-0.162 \pm 0.147) Z + (0.0241 \pm 0.0769) & \mathrm{for}~ \mathrm{BC03}, \\
		(-0.320 \pm 0.206) Z + (0.0296 \pm 0.0362) & \mathrm{for}~ \mathrm{CB10} \\
	\end{array}
\right.\end{equation}
where the 1$\sigma$ uncertainties on the regression coefficients are provided above. Large uncertainties in $\Delta ( \log \mathcal{M} ) $ for the non-solar metallicity models affect the reliability of this fit, but it is provided for reference.

\begin{figure}[h]
\begin{center}

\includegraphics[width=6.5in]{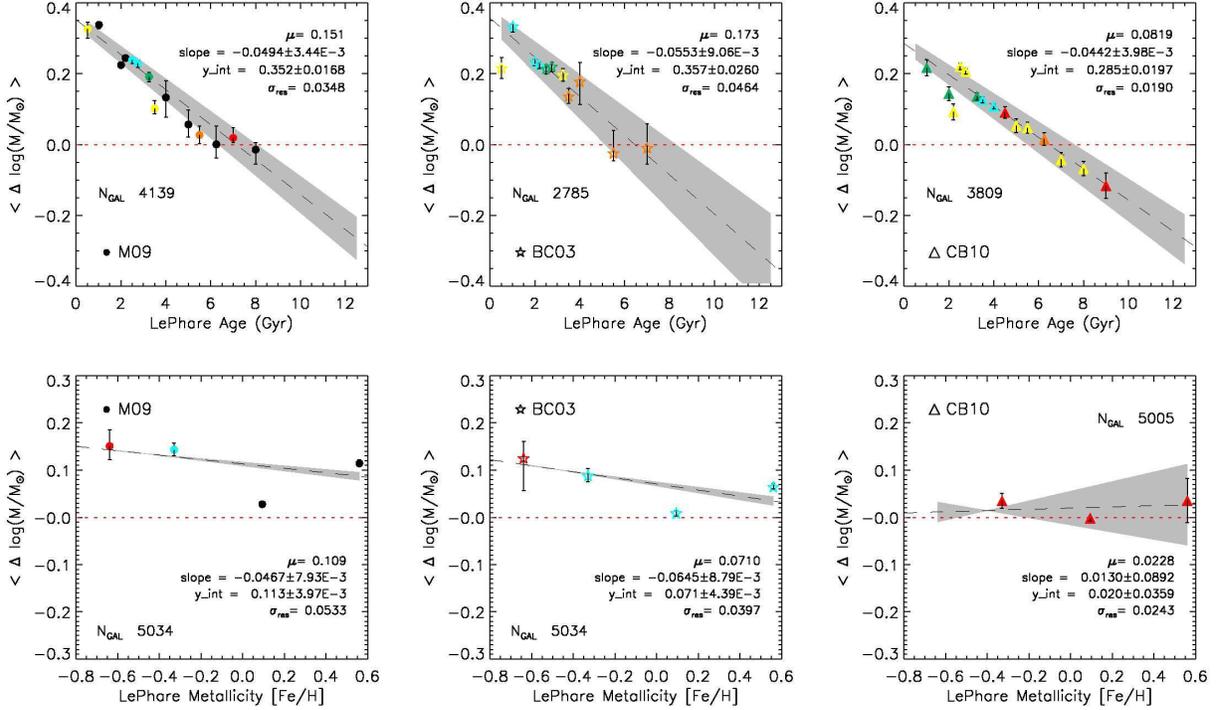}

\end{center}
\caption[XXX]
{\label{fig_mass_correction_lephare}
The \emph{mass correction}, $\epsilon$, plots as a function of photometrically determined age and metallicity, using M09 ($left$), BC03 ($center$), and CB10 ($right$) models. The color code is defined in Figure \ref{fixed_histograms_opticalNIR}. Measurements in a given bin are only considered if the bin contains $\geq$20 galaxies without catastrophic fitting failures (as returned by $LePhare$). Overplotted are linear fits to each data set. The error bars denote the 95\% CL on the median, normalized by the square root of the number of galaxies in the given bin. Provided within each plot are the mean offset ($\mu$), the standard deviation of the residuals ($\sigma_{\mathrm{res}}$), and the slope and y-intercept for a model line fit from a $\chi^{2}$-minimization.
}
\end{figure}

\begin{figure}[h]
\begin{center}

\includegraphics[width=6.5in]{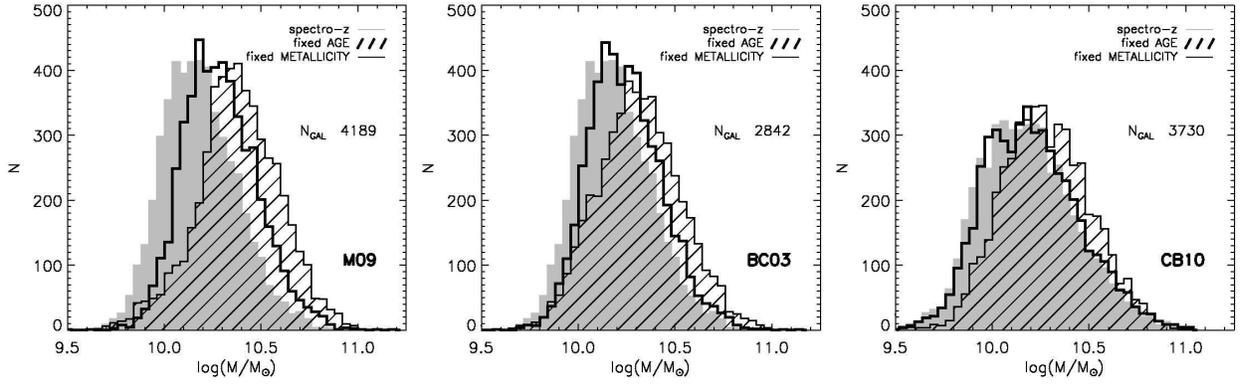}

\end{center}
\caption[XXX]
{\label{fig_mass_corrections}
Stellar mass distributions for the optical+NIR sample. Only the spectroscopic
redshift is constrained in the $gray$ histogram. The $black$ $shaded$ ($outlined$) histograms
show the stellar masses, corrected for age (metallicity) according to the
procedure described in Section \ref{sec_age}. These corrections utilize the
spectroscopic measurements from M09 ($left$), BC03 ($center$), and CB10
($right$) stellar population models, respectively.
}
\end{figure}

We can see how the distribution of stellar masses changes for our sample by applying an offset to each mass as a function of its photometrically determined metallicity, which we call the \emph{mass correction}, $\epsilon$. For galaxies with a given photometrically-determined metallicity, we add an offset calculated from the linear fit in Figure \ref{fig_mass_correction_lephare} to the stellar mass measured with only the
spectroscopic redshift constrained. Figure \ref{fig_mass_corrections} shows the
distribution of stellar masses for our optical+NIR sample, corrected for these
offsets. This technique shows that the median stellar mass changes by $\sim$0.1
dex when using the M09/BC03 models. The distribution remains more consistent in CB10 models, which may be expected given the slightly better agreement between spectroscopic and photometric metallicities using these templates (see CB10/Metallicity plot in Figure \ref{fig_starlight_v_lephare_all_delta}).

\subsubsection{Dependence on stellar age}
\label{sec_age}

As described in Section \ref{sec_ssp}, stellar masses of low redshift ETGs have
uncertainties of $\sim$0.14 dex at the 95\% CL due to the stellar
population models alone. Furthermore, the photometrically determined stellar age
is model-dependent. The relationship between dominant stellar population age and
stellar mass is not easily quantifiable, but errors in the age are propagated to
errors in the mass estimate. To this end, we have used stellar age measurements
from {\it STARLIGHT} spectral fits to fix the age of the template used in the
$\chi^{2}$-minimization.

Figure \ref{fixed_histograms_opticalNIR} shows an estimate of the systematic
bias in stellar mass for unconstrained age plotted as a function of
spectroscopically determined age. We note that when constraining the stellar age
in {\it LePhare} to its spectroscopically determined value, only 1,314 (1,810)
galaxies, using BC03 (CB10) models, have non-zero $\exp ( - \chi^{2}/2 )$
values. There is a clear trend of increasing mass difference with age.
Furthermore, using BC03 models in the SED fitting (provided spectroscopic
constraints separately from M09/BC03) yields systematically lower stellar masses
at all ages. This trend is also evident from CB10 models, though stellar masses
for unconstrained age are systematically higher below about 4 Gyr. The linear
regression fit to the optical+NIR sample yields the following functional form:

\begin{equation}
\Delta (\log \mathcal{M}) = \left\{
	\begin{array}{ll}
		(0.0359 \pm 0.0126) t + (0.0175 \pm 0.0963) & \mathrm{for}~ \mathrm{M09}, \\
		(0.0395 \pm 0.0110) t + (-0.0148 \pm 0.0837) & \mathrm{for}~ \mathrm{BC03}, \\
		(0.0467 \pm 0.0237) t + (-0.144 \pm 0.133) & \mathrm{for}~ \mathrm{CB10} \\
	\end{array}
\right.\end{equation}
where the 1$\sigma$ uncertainties on the regression coefficients are provided above. Notice that there is a significant positive trend in stellar mass underestimation with increasing age that is well
approximated by the linear model.

\begin{figure}[h]
\begin{center}

\includegraphics[width=5in]{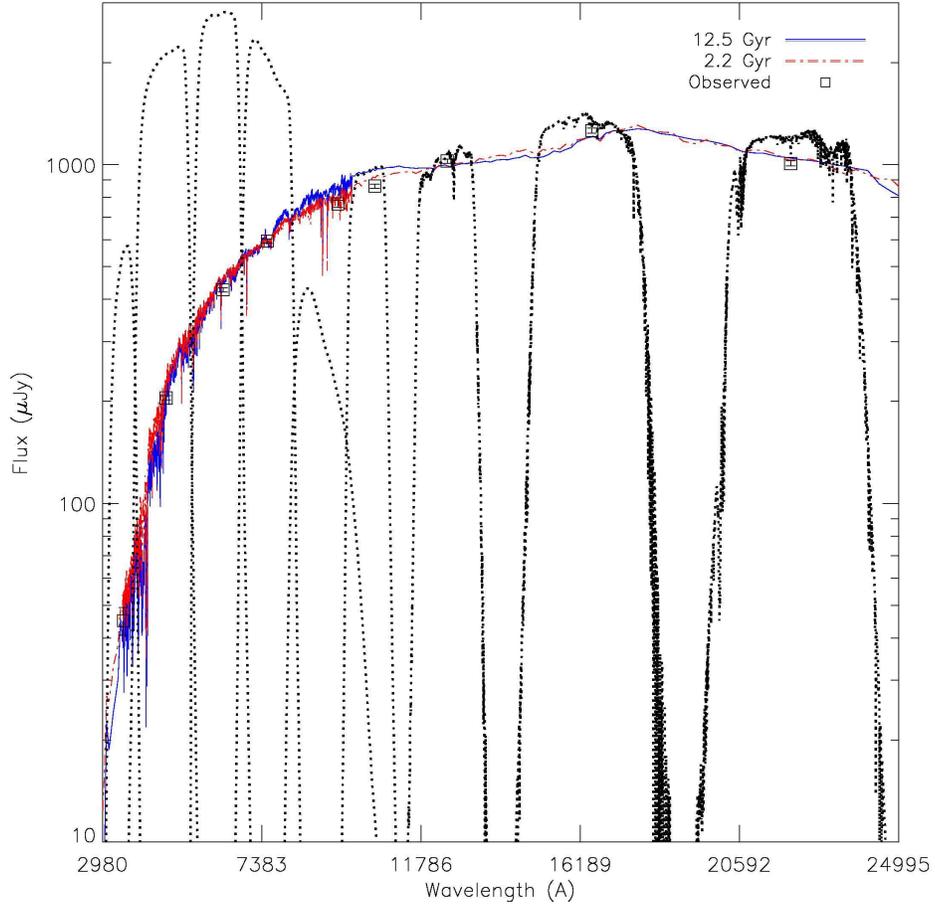}

\end{center}
\caption[XXX]
{\label{fig_sed_comparison}
Best-fit output spectra from {\it LePhare} for a single galaxy with age fixed to the spectroscopic value of 12.5 Gyr (\emph{solid blue}) and photometrically fit to 2.2 Gyr (\emph{dash-dot red}), selected from the optical+NIR sample. The observed apparent magnitudes, converted in flux, are overplotted (\emph{squares}) with their 1$\sigma$ errors. The $ugrizYJHK$ filter curves are overplotted for reference.
}
\end{figure}

This gross underestimate in mass for old stellar populations might be due to the fact
that models underestimate the ages -- possibly because of age-metallicity-extinction degeneracy or a failure of the models to reproduce the observed spectra --
and then the SED fitting code increases the luminosity scale factor accordingly. 
To illustrate this, we select a single
galaxy with spectroscopically and photometrically determined ages of 12.5 Gyr
and 2.2 Gyr, respectively, and compare the best fit redshifted, reddened
photometric SEDs for free and fixed age (see Figure \ref{fig_sed_comparison}),
where the fixed age stellar mass is higher by $\sim$0.42 dex. The galaxy is best
fit by a $Z/Z_{\sun}=0.4$ metallicity, and differs by 0.1 in $E(B-V)$, using
BC03 models. Following Equation \ref{eqn_m}, the median difference in $0.4
\times m_{\lambda^{\prime}}$ in the $g$-band is 0.06 mag (with higher flux in
the 2.2 Gyr SED) and indistinguishable from our broadband photometry alone.
Older and younger populations have very different stellar mass-to-light ratios,
where for instance the difference in $\log (\mathcal{M}_{*}/\mathcal{L}_{B})$
for the two best fit ages here is $\sim$0.67 dex (with lower $B$-band
mass-to-light ratio in the 2.2 Gyr SED). As expected, the 4000\AA~break is the
most significantly differing feature in the broadband colors of these SEDs, with
$\Delta [ \log (\mathcal{M}_{*}/\mathcal{L}_{B}) ] - \Delta (0.4 \times m_{g})$
accounting for most of the measured difference in stellar mass for this
particular object, yet it cannot be photometrically distinguished despite the high
S/N of our objects at this redshift \footnote{We compare fluxes and
mass-to-light ratios in $g$ and $B$-bands, respectively, for this object because
$\mathcal{M}_{*}/\mathcal{L}$ is provided for $B$ and $V$-bands from the BC03
code. Sloan $g$ and Johnson-Cousins $B$ filters span roughly the same wavelength
range and are just redward of the 4000\AA~break, which due to the large
difference in best fit stellar age, most constrains the continuum level of the
model spectrum.}. Furthermore, including the $K$-band photometry seems to bias
most SEDs towards young ages, assuming our ellipticals are indeed dominated by
older stars. So, we can attribute larger differences in the stellar mass due to
unknown age to the blue $\mathcal{M}_{*}/\mathcal{L}$ for different stellar
population ages. It is also true that differing spectroscopic and photometric
metallicities will produce different stellar masses, but given the trends
provided in Section \ref{sec_met}, it is likely that large differences in age
dominate the effect of age-metallicity degeneracy on the galaxy stellar mass.

We can see how the distribution of stellar masses changes for our sample by
applying an offset to each mass as a function of its photometric age, similar to the procedure described at the end of Section \ref{sec_met}. Figure \ref{fig_mass_corrections} shows the
distribution of stellar masses for our optical+NIR sample, corrected for these
offsets. This technique shows that the median stellar mass changes by as much as $\sim$0.1
dex when using the updated CB10 models. 

\section{External comparison of stellar masses}
\label{sec_mpi}
\begin{figure}[h]
\begin{center}

\includegraphics[width=4.5in]{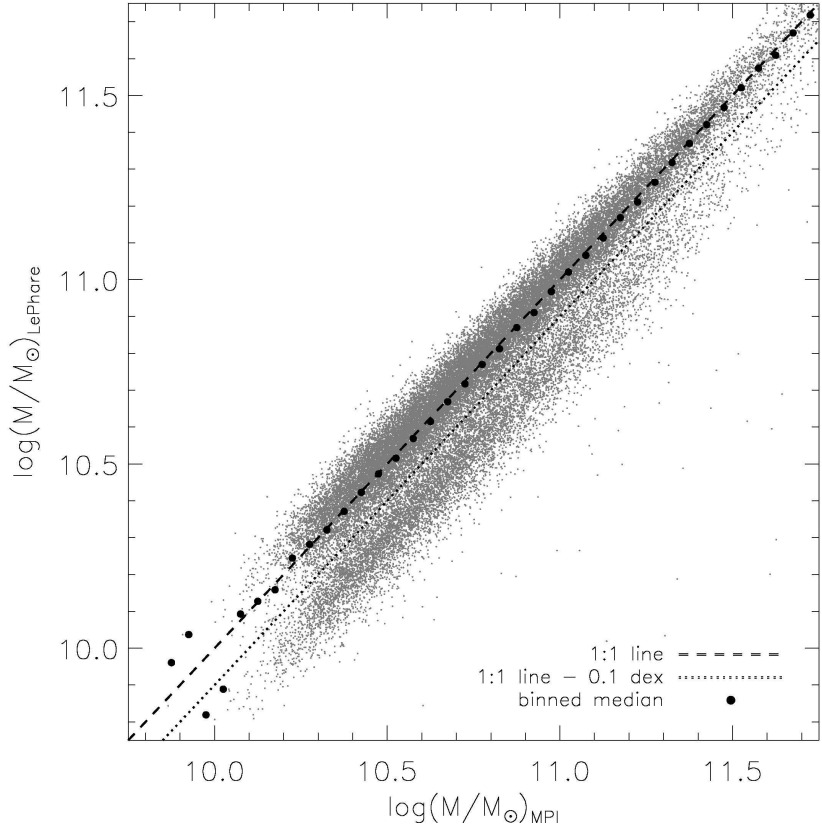}

\end{center}
\caption[XXX]
{\label{fig_MPI_vs_SPIDER}
Comparison of total stellar masses for the complete sample and the same
sample of galaxies from a group at the Max Planck Institute (obtained from
\url{http://www.mpa-garching.mpg.de/SDSS/DR7/Data/stellarmass.html}), who use
SDSS FiberMag photometry. The \emph{black} circles are obtained by median-binning the data, the \emph{dashed} line is the 1:1 line, and the \emph{dotted} line is arbitrarily drawn 0.1 dex below the 1:1 line to separate the two trends observed here.
}
\end{figure}

\begin{figure}[h]
\begin{center}

\includegraphics[width=6.5in]{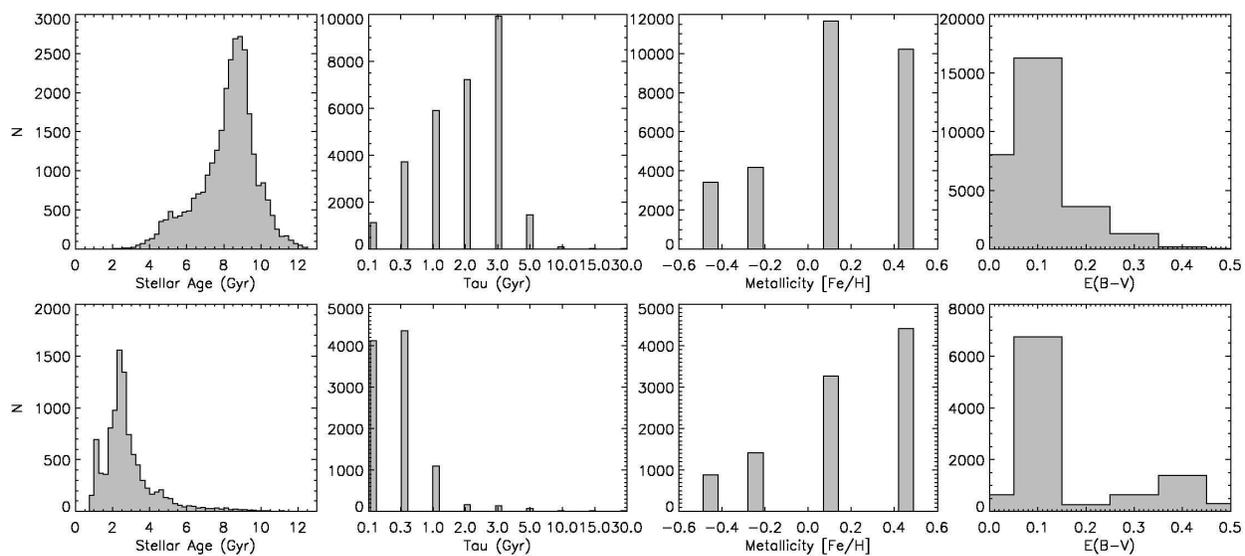}

\end{center}
\caption[XXX]
{\label{fig_hist_2trend}
Distributions of age, star-formation decay time scale, metallicity, and extinction for galaxies lying above (\emph{top} panels) and below (\emph{bottom} panels) the \emph{dotted} line in Figure \ref{fig_MPI_vs_SPIDER}.
}
\end{figure}

As a comparison to related work, we plot stellar masses obtained from our
SED fits against those derived from a group at the Max Planck Institute (see
Figure \ref{fig_MPI_vs_SPIDER}), using the complete sample and BC03 stellar
population models \footnote{However, detailed parameters, such as IMF, star
formation history, extinction law, etc. are unspecified on their website (see
\url{http://www.mpa-garching.mpg.de/SDSS/DR7/mass_comp.html}), with differences
likely contributing to the observed scatter.}. We convert to total stellar mass
by dividing each set of measured stellar masses by an aperture correction,
corresponding to the fraction of $z$-band light contained within the $i$-band
Kron aperture ({\it LePhare}) and the SDSS fiber ({\it MPI}). This procedure assumes
that the stellar mass in a galaxy is distributed in the same way as the $z$-band light.
We notice two trends in this comparison, with $\sim$75\% of the galaxies lying above an arbitrary line drawn 0.1 dex below the 1:1 line. The subset of galaxies that lie above this line are in excellent agreement with those from MPI, with a negligible offset and 2$\sigma$ uncertainty of $\sim$0.1 dex, likely due in part to the difference between how Kron and SDSS FiberMags are measured (e.g. correcting the latter for emission lines) as well as potential differences in model parameters. The distributions of best fit age, star-formation decay time scale, metallicity, and extinction are plotted in Figure \ref{fig_hist_2trend} for galaxies lying above (\emph{top} panels) and below (\emph{bottom} panels) the \emph{dotted} line drawn in Figure \ref{fig_MPI_vs_SPIDER}. We see that with metallicity and extinction being similar for the two samples, the contribution from a high $t/\tau$ is a redder galaxy -- that is, such a galaxy has enough e-folding times to allow the younger stars to fade. This produces a degeneracy in the shape of the SEDs, as compared to a similar $t/\tau$ for an older galaxy, resulting in a lower mass for the younger SED. We have directly verified this degeneracy for several objects with similar $t/\tau$, metallicity, extinction, and redshift, but different ages.

We also compared the stellar masses that we present here with those presented in Paper IV, finding trends similar to those described above. Specifically, we find a negligible offset (0.003 dex) with a modest 2$\sigma$ scatter of $\sim$0.1 dex for galaxies (65\%) lying above a line, drawn similarly to the \emph{dotted} line in Figure \ref{fig_MPI_vs_SPIDER}. The stellar population model differences between the stellar masses presented in this paper -- i.e. for BC03 models, with a Chabrier IMF, Cardelli extinction law, and spectroscopic redshift -- and those presented in Paper IV is the permitted model decay time scales for star-formation ($\tau$), where in this paper we also include $\tau=0.1, 0.3, \mathrm{and}~30.0$ Gyr models, and use of the Calzetti law in Paper IV (which induces negligible offset in the stellar mass). Another difference is that in this paper we use all available bands to scale the SED, so that the scale factor is less sensitive to errors in any one given band. In Paper IV, we used only $K$-band to scale the SED, due to its supposed independence from the effects of dust extinction and young, luminous stars (see Section \ref{sec_41}).

\section{Summary \& conclusions}

We measured systematic errors in galaxy stellar masses due to different
ingredients in widely used models from stellar population synthesis techniques.
Using a large sample of high S/N, low redshift ETGs, we want to accomplish two
primary goals

-- Determine at a high-level of significance, the statistical biases and
uncertainties inherent in a stellar mass estimate;

-- Provide robust corrections to a sample of photometrically determined galaxy
stellar masses, when spectroscopy is unavailable.

We first presented a comparison of BC03, CB10, and PEGASE.2 stellar mass
estimates, finding that masses have dispersions of $\sim$$0.10-0.15$ dex at the
95\% CL, with this uncertainty arguably being due to the different treatment of
the TP-AGB stellar evolutionary phase. In general, our tests find no statistically significant
systematic mass bias at the $2\sigma$ level due to variations of the stellar population models, extinction
laws, and photometric bandpasses. This includes varying scaling bandpasses in the fitting technique as
well as the magnitudes used in generating the photometric SED. However, a systematic offset of about $-0.23$ dex (inconsistent with zero at the $2\sigma$ level) is observed between stellar masses predicted
from Chabrier and Salpeter IMFs. The newer TP-AGB calculations by \cite{2007A&A...469..239M} affecting
$K$-band magnitudes also improve the quality of $u$-band fits, where the dispersion in
galaxy stellar mass measured with and without $u$-band data is improved by
nearly 20\% from BC03 to CB10 models. It is interesting to note that
stellar mass estimates are more consistent with no near-infrared photometry than
they are without $u$-band data, at least using BC03 models with this sample of
ETGs.


The observed systematic stellar mass biases and dispersions for our optical+NIR
and complete sample of ETGs are summarized in Table \ref{tab_summary}. As
expected for this redshift range, using the photo-z produces the largest
systematic uncertainties in the stellar mass estimate. We find that a
\cite{Chabrier} IMF produces lower stellar masses than a \cite{Salpeter} IMF by
about 0.227 (0.225) dex for BC03 (CB10) models, in agreement with other
studies, but lower by about 0.1 dex than the difference predicted by direct
integration of the formulae over the mass range 0.1$-$100~$\mathcal{M}_{\sun}$ (Appendix~\ref{app:imf}). For the optical+NIR sample, this difference falls within the 3$\sigma$ confidence limits of the observational estimate (i.e. 0.23 +0.15/-0.10 for BC03), while for the complete sample, it is within the 4$\sigma$ confidence limits for BC03 and highly inconsistent with zero ($>4\sigma$ deviation) for CB10.

\begin{figure}[h]
\begin{center}

\includegraphics[width=6.5in]{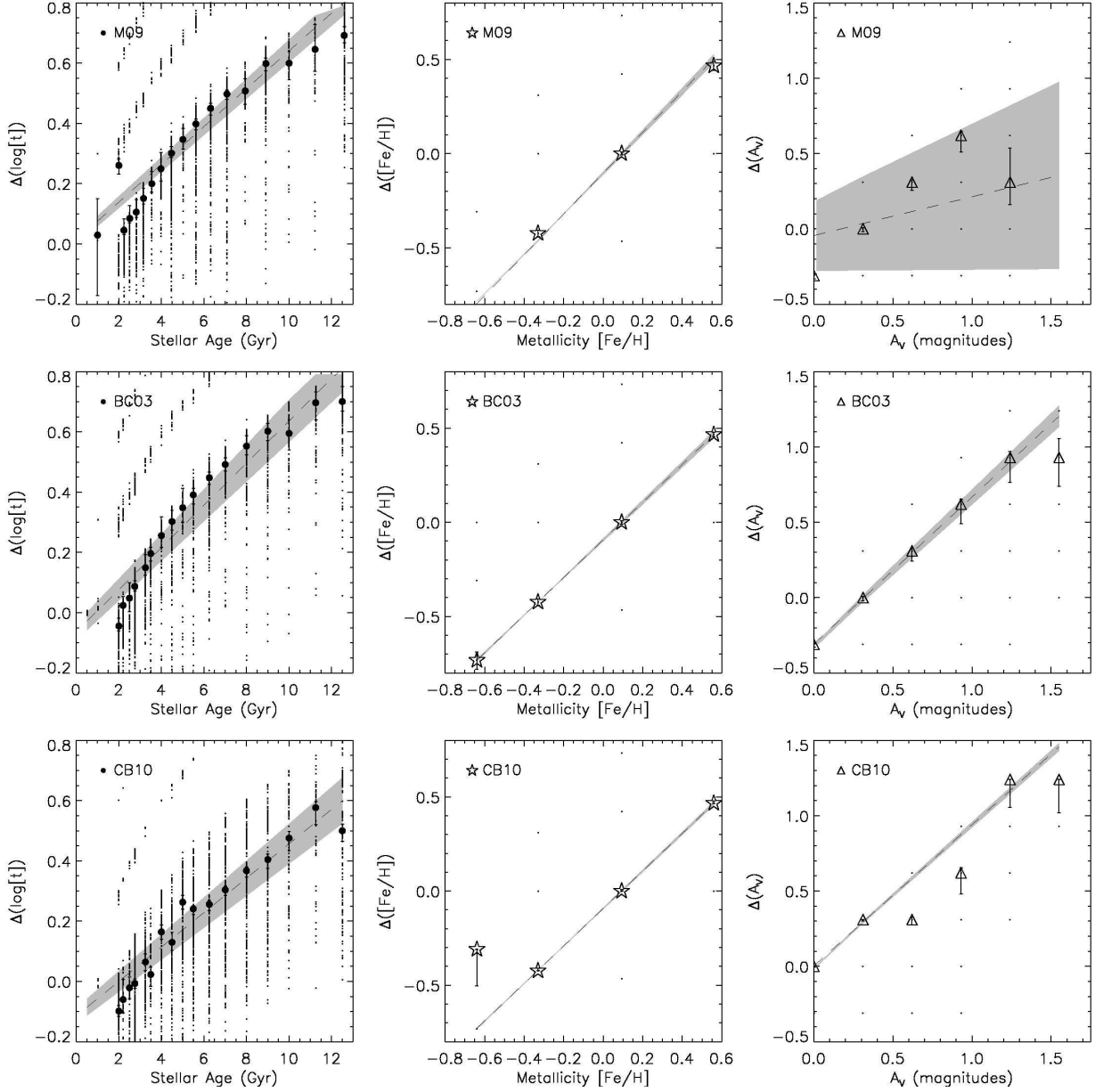}

\end{center}
\caption[XXX]
{\label{fig_starlight_v_lephare_all_delta}
Each plot shows differences ($STARLIGHT-LePhare$) in spectroscopic and photometric parameters $t$, $Z$, and $A_{V}$, as a function of their spectroscopic values,
measured with {\it STARLIGHT}. $Black$ symbols represent median values in each bin. Linear
fits ($dashed$ lines) to the data are presented, with their 1-sigma uncertainties in the $gray$ shaded region.
}
\end{figure}

An important part of this work is the investigation of systematic effects on the
photometrically determined stellar mass when spectroscopic constraints on age,
metallicity, and extinction are unavailable, and to provide statistically robust
stellar mass corrections as a function of these photometric parameters for low
redshift early-type galaxies. We used 5,068 objects with $ugrizYJHK$ photometry
and spectroscopy to achieve this goal. Offsets from extinction constraints are
the smallest of these systematic effects; this implies that the SED fitting is less
sensitive to the coarse extinction grid employed by stellar population models when estimating the stellar mass. We find a negligible trend with $A_{V}$ in stellar
mass difference for the optical+NIR sample. However, there is about a 0.15 dex
bias towards higher masses for all extinctions when $A_{V}$ is unknown, using
M09/CB10 models. The dispersion in these mass differences
is about $\pm0.5$ dex (consistent with zero), slightly lower for BC03 models, yet larger than
uncertainties introduced by unknown age/metallicity, mainly because of the
smaller number of galaxies in a given extinction bin. Fitting broadband
photometry seems to produce lower interstellar extinction than spectral fitting
for our sample of ETGs, as shown in the {\it right} panels in Figure \ref{fig_starlight_v_lephare_all_delta}.
This would lower the $0.4 \times m_{\lambda^{\prime}}$ term in Equation
\ref{eqn_m} and produce a higher stellar mass estimate. We comment
that for general surveys, it might be more appropriate to fit an SED with one extinction law \citep[e.g.][]{Calzetti}, and then refit with a more appropriate law once an approximate age is established.

The positive trend in $\Delta ( \log \mathcal{M} )$ with increasing metallicity
is simply a result of the inverse trend shown in the {\it middle} panels of Figure
\ref{fig_starlight_v_lephare_all_delta}. That is, spectral fits on average produce
lower metallicities than broadband photometric SED fitting for these objects;
and for any given age, a lower metallicity SSP has a lower mass. This is merely a  reflection of
the failure of such a coarse metallicity grid to constrain the model in the
SED fitting, as expected. \cite{1987AJ.....94..899D} show that
the size of the 4000\AA~break in spheroidal galaxies is quite insensitive to
metallicity, so even provided a wider range of stellar population metallicities,
our result should remain unchanged in this regard for (old) ETGs. We chose a
larger grid of stellar ages for M09/BC03/CB10 models, ranging from $0.5-12.5$
Gyr to constrain age in the $\chi^{2}$-minimization. We measure remarkably lower
stellar masses when age is unknown for all young/old populations (except
$\lesssim4$ Gyr for CB10 models), with uncertainties of around $0.2-0.3$ dex.
Indeed, for our low redshift ETGs with older stellar populations, we would be
underestimating the stellar mass content by a factor of 2 (for galaxies with
$t\sim10$ Gyr from BC03/CB10 models). We can argue that the increasing trend with stellar age in Figure \ref{fixed_histograms_opticalNIR} is directly due to the overall underestimate in photometrically determined age for all models (see {\it left} panels of Figure \ref{fig_starlight_v_lephare_all_delta}), whereas we argue in Section \ref{sec_age}, a difference of $\sim$0.7 dex in the best fit age can yield a stellar mass difference of $\sim$0.4 dex, explained by
varying $\mathcal{M}_{*}/\mathcal{L}$ ratios between different stellar age SEDs. Notice that unknown
ages, resulting from a mismatch between the true star formation history of a galaxy and the
simplified form used in SED fits (e.g. an exponentially declining SFR), can also affect significantly the
stellar mass determination of high-redshift  galaxies, as discussed, e.g., by \cite{2007A&A...474..443P} and \cite{2010ApJ...725.1644L}.

In conclusion, we find uncertainties in the galaxy stellar mass due to different
stellar population models, IMFs, and extinction laws that are much less than
uncertainties introduced by using the photo-z over this small redshift range.
More notable are offsets measured between Chabrier/Salpeter IMFs,
Cardelli/CF2000 extinction laws, and changes in TP-AGB evolutionary
prescriptions -- between CB10/PEGASE.2 models. The discrepancies in ages,
metallicities, and extinctions between spectroscopic and photometric measurement
techniques propagate to uncertainties in the stellar mass estimates, yielding
notably lower masses when stellar age is unknown. Given the coarse grid of
stellar metallicities, the age-metallicity degeneracy is likely to contribute to the
large discrepancies in spectroscopic and photometric measurements. Finally, we
proposed a \emph{mass correction} to our sample that incorporates these
systematic offsets as a function of their photometrically determined age and
metallicity, finding that the sample median mass increases by a factor of
roughly 1.3. We would like to comment that these results apply to optical and
near-infrared broadband photometry of low redshift ETGs. Detailed quantification
of the systematic errors involved in SED fitting have produced results
consistent with what other studies have found and have allowed us to obtain
consistent stellar mass estimates for this sample.

We have used data from the 4th data release of the UKIDSS survey, which
is described in detail in \citet{War07}. The UKIDSS project is
defined in~\citet{Law07}. UKIDSS uses the UKIRT Wide Field Camera
(WFCAM; \cite{Casali07}). The photometric system is described in
\cite{Hewett}, and the calibration is described in \cite{Hodgkin09}. The pipeline processing and science archive are described in Irwin et al. (2011, in prep) and \cite{Hambly08}. Funding for
the SDSS and SDSS-II has been provided by the Alfred P. Sloan
Foundation, the Participating Institutions, the National Science
Foundation, the U.S. Department of Energy, the National Aeronautics
and Space Administration, the Japanese Monbukagakusho, the Max Planck
Society, and the Higher Education Funding Council for England. The
SDSS Web Site is \url{http://www.sdss.org}. The SDSS is managed by the
Astrophysical Research Consortium for the Participating Institutions.
The Participating Institutions  are the American Museum of Natural
History, Astrophysical Institute Potsdam, University of Basel,
University of Cambridge, Case Western Reserve University, University
of Chicago, Drexel University, Fermilab, the Institute for Advanced
Study, the Japan Participation Group, Johns Hopkins University, the
Joint Institute for Nuclear Astrophysics, the Kavli Institute for
Particle Astrophysics and Cosmology, the Korean Scientist Group, the
Chinese Academy of Sciences (LAMOST), Los Alamos National Laboratory,
the Max-Planck-Institute for Astronomy (MPIA), the
Max-Planck-Institute for Astrophysics (MPA), New Mexico State
University, Ohio State University, University of Pittsburgh,
University of Portsmouth, Princeton University, the United States
Naval Observatory, and the University of Washington. The authors would also like to thank
Olivier Ilbert and Stephane Arnouts for help with using $LePhare$.

\begin{sidewaystable}
\centering
\caption{Summary of results: \emph{optical+NIR}. \emph{Offset} is the median of the $\Delta ( \log
\mathcal{M} )$ histograms and $2\sigma$ is the $\pm$2-sigma uncertainty on the
median. \emph{Offset} is provided for the fixed$-$free parameter tests (Figure
\ref{fixed_histograms_opticalNIR}) at the median spectroscopic values of $t$, Z,
and $A_{V}$ for BC03 and CB10 models. Unless otherwise noted, these tests assume
a spectroscopic redshift, Chabrier IMF, and Cardelli extinction law. Note that galaxies fit with the median spectroscopic age of 4.0 Gyr (CB10) suffered catastrophic fitting failures, and so the offset and uncertainties are reported here for an age of 4.5 Gyr.}
\begin{center}
\begin{tabular}{cccccccc}

\hline
\hline
& \multicolumn{2}{c}{BC03} & \multicolumn{2}{c}{CB10} \\
& Offset & $2\sigma$ & Offset & $2\sigma$ & Note \\
\hline
spectro-z $-$ photo-z & 0.0128 & +0.804/-0.500 & 0.0555 & +0.771/-0.524 &
$0.000 \leq z_{p} \leq 0.300$, $9 < \log \mathcal{M}/\mathcal{M}_{\sun} < 12$ \\
BC03 $-$ CB10 & -4.64E-3 & +0.118/-0.163 & -4.64E-3 & +0.118/-0.163 & STELIB
spectral library \\
BC03 (CB10) $-$ PEGASE.2 & 0.0962 & +0.0837/-0.112 & 0.0858 &
+0.0934/-0.132 & --- \\
\hline
Chabrier $-$ Salpeter & -0.227 & +0.0855/-0.0323 & -0.225 &
+0.0441/-0.0281 & STELIB spectral library \\
\hline
Cardelli $-$ Calzetti & -4.50E-3 & +0.0711/-0.140 & -1.00E-4 &
+0.109/-0.0375 & Cardelli/Calzetti from {\it LePhare} code \\
Cardelli $-$ CF2000 & -0.104 & +0.142/-0.170 & -0.0680 & +0.156/-0.131 &
$\tau_{V}=1.0$, $\mu=0.3$ (in GALAXEV code) \\
\hline
$u-$ no-$u$ & -0.0257 & +0.131/-0.151 & -6.30E-3 & +0.115/-0.121 & --- \\
$ugrizYJHK-ugriz$ & -0.0566 & +0.0868/-0.125 & -0.0419 & +0.147/-0.108 &
--- \\
\hline
Fixed$-$Free & & & \\
$t$ & 0.0833 & +0.108/-0.144 & 0.0590 & +0.284/-0.201 &
$t_{med}=3.25~(4.50)$ Gyr, BC03 (CB10) \\
$Z$ & 0.0587 & +0.191/-0.0442 & -1.50E-3 & +0.0273/-0.151 &
[Fe/H]$_{med}=0.0932~(0.0932)$, BC03 (CB10) \\
$A_{V}$ & 0.0276 & +0.280/-0.147 & -0.164 & +0.554/-0.488 &
$A_{V,med}=0.12~(0.10)$, BC03 (CB10) \\
\hline

\end{tabular}
\end{center}
\label{tab_summary}
\end{sidewaystable}

\clearpage

\begin{sidewaystable}
\centering
\caption{Summary of results: \emph{complete}. \emph{Offset} is the median of the $\Delta ( \log
\mathcal{M} )$ histograms and $2\sigma$ is the $\pm$2-sigma uncertainty on the
median. \emph{Offset} is provided for the fixed$-$free parameter tests (Figure
\ref{fixed_histograms_opticalNIR}) at the median spectroscopic values of $t$, Z,
and $A_{V}$ for BC03 and CB10 models. Unless otherwise noted, these tests assume
a spectroscopic redshift, Chabrier IMF, and Cardelli extinction law. Again, the reader is cautioned on use of the results in this table, per the reasons outlined in Section \ref{sec_results}.}
\begin{center}
\begin{tabular}{cccccccc}

\hline
\hline
& \multicolumn{2}{c}{BC03} & \multicolumn{2}{c}{CB10} \\
& Offset & $2\sigma$ & Offset & $2\sigma$ & Note \\
\hline
spectro-z $-$ photo-z & 0.0281 & +0.780/-0.330 & 0.0185 & +0.772/-0.316 &
$0.000 \leq z_{p} \leq 0.300$, $9 < \log \mathcal{M}/\mathcal{M}_{\sun} < 12$ \\
BC03 $-$ CB10 & 0.0131 & +0.0424/-0.0326 & 0.0131 & +0.0424/-0.0326 & STELIB
spectral library \\
BC03 (CB10) $-$ PEGASE.2 & 0.0871 & +0.0516/-0.0908 & 0.0802 &
+0.0550/-0.151 & --- \\
\hline
Chabrier $-$ Salpeter & -0.233 & +0.0144/-0.0140 & -0.231 &
+0.0173/-0.0151 & STELIB spectral library \\
\hline
Cardelli $-$ Calzetti & -0.0193 & +0.0777/-0.0386 & 5.10E-3 &
+0.0578/-0.0718 & Cardelli/Calzetti from {\it LePhare} code \\
Cardelli $-$ CF2000 & -0.0693 & +0.0918/-0.0501 & -0.0225 & +0.0941/-0.0433 &
$\tau_{V}=1.0$, $\mu=0.3$ (in GALAXEV code) \\
\hline
$u-$ no-$u$ & -0.0128 & +0.0410/-0.102 & -0.0166 & +0.0571/-0.187 & --- \\
\hline

\end{tabular}
\end{center}
\label{tab_summary_complete}
\end{sidewaystable}

\clearpage

\appendix

\section{SDSS fiber vs. Kron radius}
\label{app:fiber_kron_mags}
\begin{figure}[h]
\begin{center}

\includegraphics[width=6.5in]{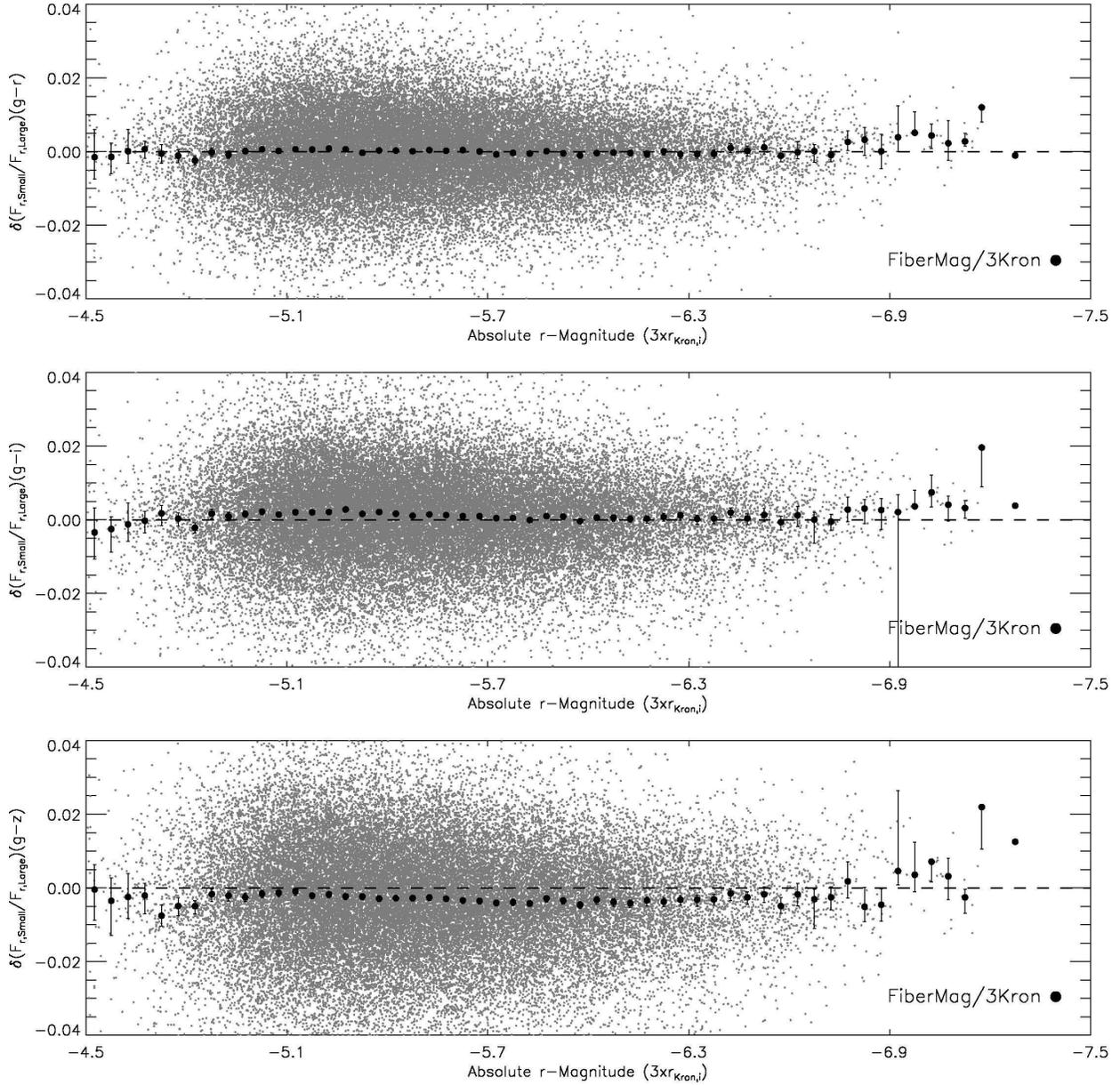}

\end{center}
\caption[XXX]
{\label{fig_fiber_v_kron_delta}
Difference in flux ratio of SDSS FiberMags over Kron magnitudes between the 
$g$ and $x$ wavebands, with $x=[riz]$, as a function of the absolute $r$-band magnitude
 for the complete sample of ETGs. \emph{Black} circles are obtained by median-binning the data,
with error bars marking the 2$\sigma$ uncertainty on median values. Notice that 
there is no variation in the difference of flux ratios from $g$ through $z$,
i.e. no significant color gradient between the Fiber and Kron apertures.
In $g-z$ there is a small negative offset, but it is less than 0.005 magnitudes.
}
\end{figure}

In this paper, we provide spectroscopic parameters measured within the SDSS
fiber radius of $1.5^{\prime\prime}$. However, photometric magnitudes are
measured using an adaptive aperture radius of $3 \times r_{K,i}$ (see Section
\ref{sec_sample}). Figure \ref{fig_fiber_v_kron_delta} shows the difference 
in flux ratio of SDSS FiberMags to Kron magnitudes between the wavebands $g$
and $x$, with $x=riz$, as a function of $M_{r}$ for the
complete sample. The plot shows that there is essentially no color gradient
between the $1.5^{\prime\prime}$ and $3 \times r_{K,i}$ apertures, from $g$ through 
$z$. Indeed, the FiberMag/($3 \times r_{K,i}$) flux ratios have color consistent with zero inside
the $2\sigma$ error bars for most of the absolute magnitude range, with no
noticeable dependence on $M_{r}$. We conclude that it is reasonable to compare
{\it STARLIGHT} spectroscopic parameters, measured within the fiber aperture, to
those from SED fitting, measured within the $3 \times r_{K,i}$ aperture.

\section{The initial mass function (IMF)}
\label{app:imf}

\begin{figure}[h]
\begin{center}

\includegraphics[width=6.5in]{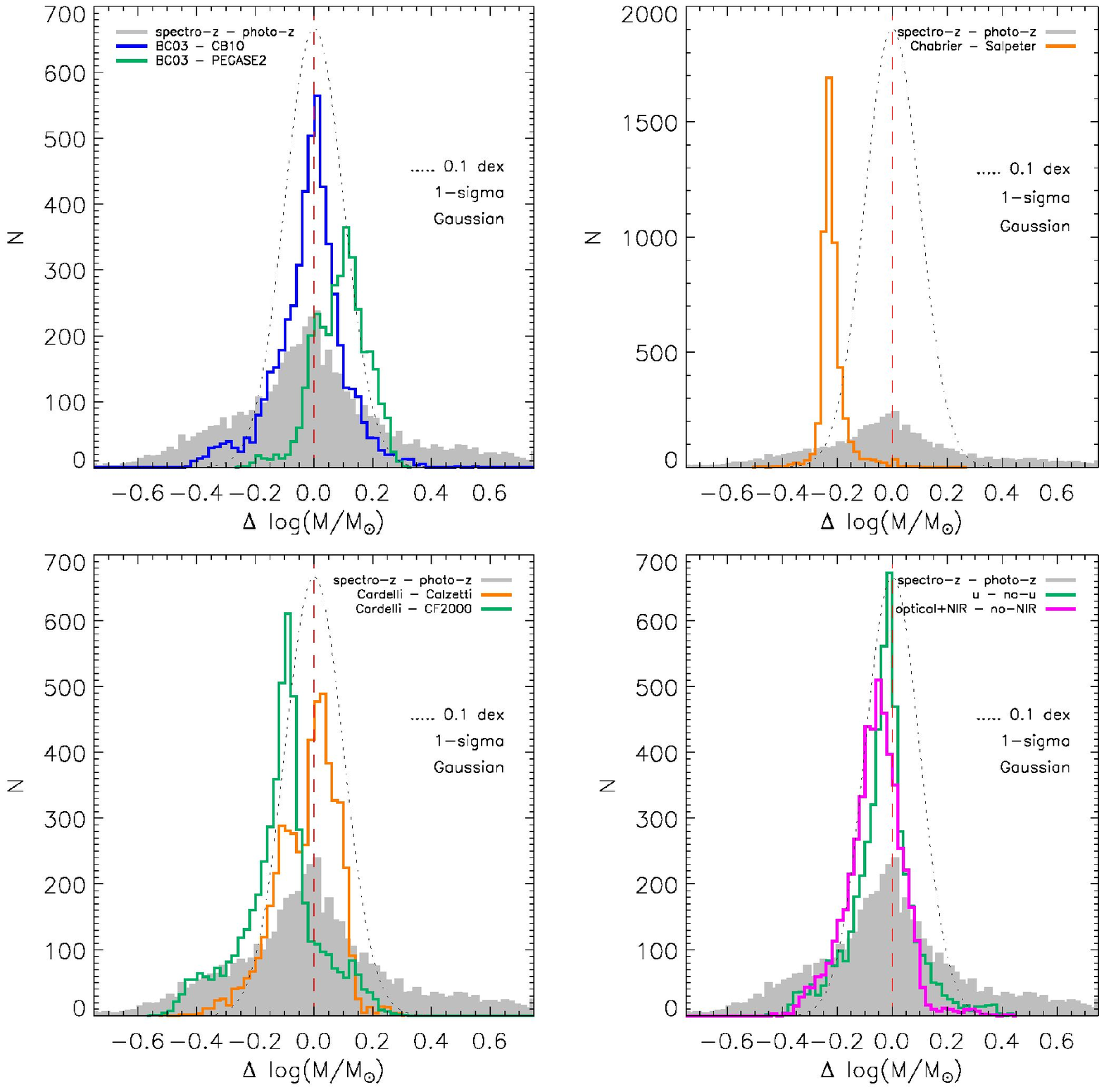}

\end{center}
\caption[XXX]
{\label{fig_histogram_masses_BC03}
Histograms of $\Delta ( \log \mathcal{M} )$, defined in Section
\ref{sec_method}, for the optical+NIR sample and various BC03 model parameters.
With the exception of the spectro-z$-$photo-z mass plot (\emph{gray}),
the redshift is fixed in each case. A Chabrier IMF and Cardelli extinction law
is assumed, unless otherwise noted. Overplotted are a vertical $dashed$
line and a Gaussian with $\sigma=0.1$ dex centered on zero.
}
\end{figure}

\begin{figure}[h]
\begin{center}

\includegraphics[width=6.5in]{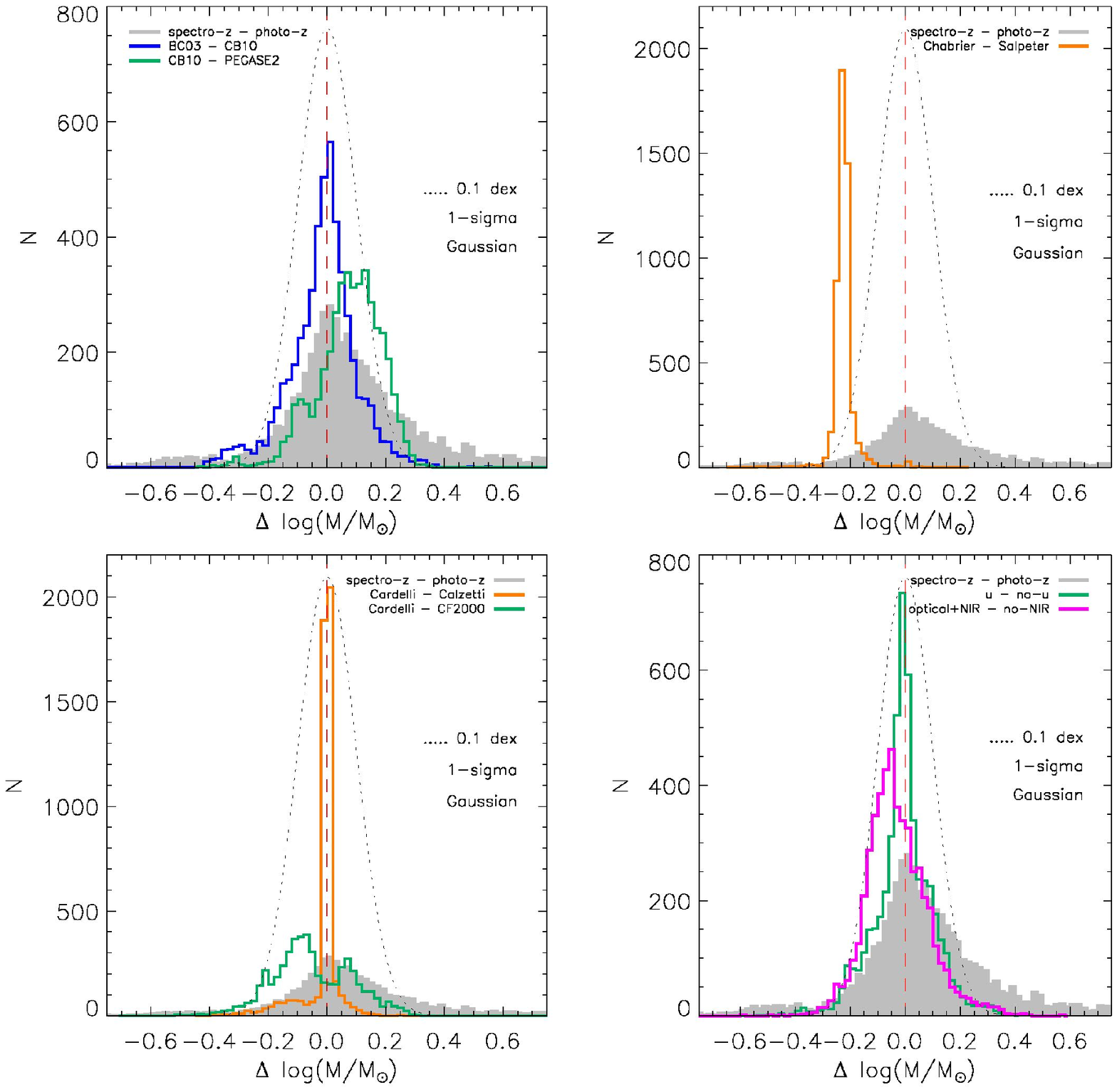}

\end{center}
\caption[XXX]
{\label{fig_histogram_masses_CB10}
Histograms of $\Delta ( \log \mathcal{M} )$, defined in Section
\ref{sec_method}, for the optical+NIR sample and various CB10 model parameters.
With the exception of the spectro-z$-$photo-z mass plot (\emph{gray}),
the redshift is fixed in each case. A Chabrier IMF and Cardelli extinction law
is assumed, unless otherwise noted. Overplotted are a vertical $dashed$ line
and a Gaussian with $\sigma=0.1$ dex centered on zero. A color version of this figure is available online.
}
\end{figure}

Expressions for the initial distribution of stars in a stellar population as a
function of mass, known as the stellar initial mass function (IMF), have been
proposed by several authors (e.g. \citealp{Salpeter, Scalo86, Chabrier}). In
this paper, we primarily use the Chabrier IMF for BC03/CB10, parameterized as

\begin{equation}
\label{eqn_chab}
\xi ( m ) \propto \left\{
	\begin{array}{ll}
		\exp [ -\frac{ ( \log m - \log m{_c} ) ^{2}}{2\sigma^{2}} ] & ,
\mathrm{for}~ 0.1 \leq m/\mathcal{M}_{\sun} \leq 1, \\
		m^{-2.3} & , \mathrm{for}~ 1 < m/\mathcal{M}_{\sun} \leq 100, \\
	\end{array}
\right.\end{equation} \\
with $m_{c}=0.08\mathcal{M}_{\sun}$ and $\sigma=0.69$, and the Scalo IMF for
PEGASE.2, approximated here by two power-law segments:

\begin{equation}
\label{eqn_scalo}
\xi ( m ) \propto \left\{
	\begin{array}{ll}
		m^{-1.53} & , \mathrm{for}~ 0.1 \leq m/\mathcal{M}_{\sun} \leq
1, \\
		m^{-2.67} & , \mathrm{for}~ 1 < m/\mathcal{M}_{\sun} < 120. \\
	\end{array}
\right.\end{equation} \\
The logarithmic slope of $\xi(m) \propto m^{-2.35}$ (B3) for the Salpeter IMF is
tested against the Chabrier form in Section \ref{sec_ssp}. We can compare the
results to the analytical approximations, derived by integrating Equations (B1),
(B2), and (B3) as $m \times \xi(m)$. Salpeter masses are expected to be a factor 
of 2.15 ($\sim$0.33~dex) larger
than Chabrier, and Scalo masses are expected to be larger by a factor of 1.05 over Chabrier.
The analytical result ($0.33$~dex) falls within the 3$\sigma$ confidence limits of the Chabrier/Salpeter
comparison in Figures \ref{fig_histogram_masses_BC03} and \ref{fig_histogram_masses_CB10}. We then take the difference in stellar mass
between the Chabrier and Scalo IMFs computed above and apply this correction to
$\Delta (\log \mathcal{M})$ for BC03(CB10)$-$PEGASE.2, as described in the text.

\clearpage

\begin{table}[htdp]
\caption{Stellar population model parameters (BC03/CB10/PEGASE.2).
}
\begin{center}
\begin{tabular}{c|cc|c}

\hline
\hline
& \multicolumn{2}{|c|}{BC03/CB10} & PEGASE.2 \\
\hline
& $\tau$ & & \\
Number & (Gyr) & $Z$ & SFR \\
\hline
1 & 0.1 & $0.2 \times Z_{\sun}$ & $\delta(t=0)$ \\
2 & 0.1 & $0.4 \times Z_{\sun}$ & $\tau=0.1$ \\
3 & 0.1 & $Z_{\sun}$ & $\tau=0.3$ \\
4 & 0.1 & $3 \times Z_{\sun}$ & $\tau=0.5$ \\
5 & 0.3 & $0.2 \times Z_{\sun}$ & $\tau=0.7$ \\
6 & 0.3 & $0.4 \times Z_{\sun}$ & $\tau=1.0$ \\
7 & 0.3 & $Z_{\sun}$ & $\tau=2.0$ \\
8 & 0.3 & $3 \times Z_{\sun}$ & $\tau=3.0$ \\
9 & 1.0 & $0.2 \times Z_{\sun}$ & $\tau=5.0$ \\
10 & 1.0 & $0.4 \times Z_{\sun}$ & $\tau=7.0$ \\
11 & 1.0 & $Z_{\sun}$ & $\tau=9.0$ \\
12 & 1.0 & $3 \times Z_{\sun}$ & $\tau=10.0$ \\
13 & 2.0 & $0.2 \times Z_{\sun}$ & $\tau=15.0$ \\
14 & 2.0 & $0.4 \times Z_{\sun}$ & $\tau=20.0$ \\
15 & 2.0 & $Z_{\sun}$ & $0.5 \times 10^{-4} \mathcal{M}_{\sun} Myr^{-1}$ \\
 & & & for $t \leq 20$ Gyr \\
16 & 2.0 & $3 \times Z_{\sun}$ & --- \\
17 & 3.0 & $0.2 \times Z_{\sun}$ & --- \\
18 & 3.0 & $0.4 \times Z_{\sun}$ & --- \\
19 & 3.0 & $Z_{\sun}$ & --- \\
20 & 3.0 & $3 \times Z_{\sun}$ & --- \\
21 & 5.0 & $0.2 \times Z_{\sun}$ & --- \\
22 & 5.0 & $0.4 \times Z_{\sun}$ & --- \\
23 & 5.0 & $Z_{\sun}$ & --- \\
24 & 5.0 & $3 \times Z_{\sun}$ & --- \\
25 & 10.0 & $0.2 \times Z_{\sun}$ & --- \\
26 & 10.0 & $0.4 \times Z_{\sun}$ & --- \\
27 & 10.0 & $Z_{\sun}$ & --- \\
28 & 10.0 & $3 \times Z_{\sun}$ & --- \\
29 & 15.0 & $0.2 \times Z_{\sun}$ & --- \\
30 & 15.0 & $0.4 \times Z_{\sun}$ & --- \\
31 & 15.0 & $Z_{\sun}$ & --- \\
32 & 15.0 & $3 \times Z_{\sun}$ & --- \\
33 & 30.0 & $0.2 \times Z_{\sun}$ & --- \\
34 & 30.0 & $0.4 \times Z_{\sun}$ & --- \\
35 & 30.0 & $Z_{\sun}$ & --- \\
36 & 30.0 & $3 \times Z_{\sun}$ & --- \\
\hline
\end{tabular}
\end{center}
\label{tab_models}
\end{table}

\begin{figure}[h]
\begin{center}

\includegraphics[width=6.5in]{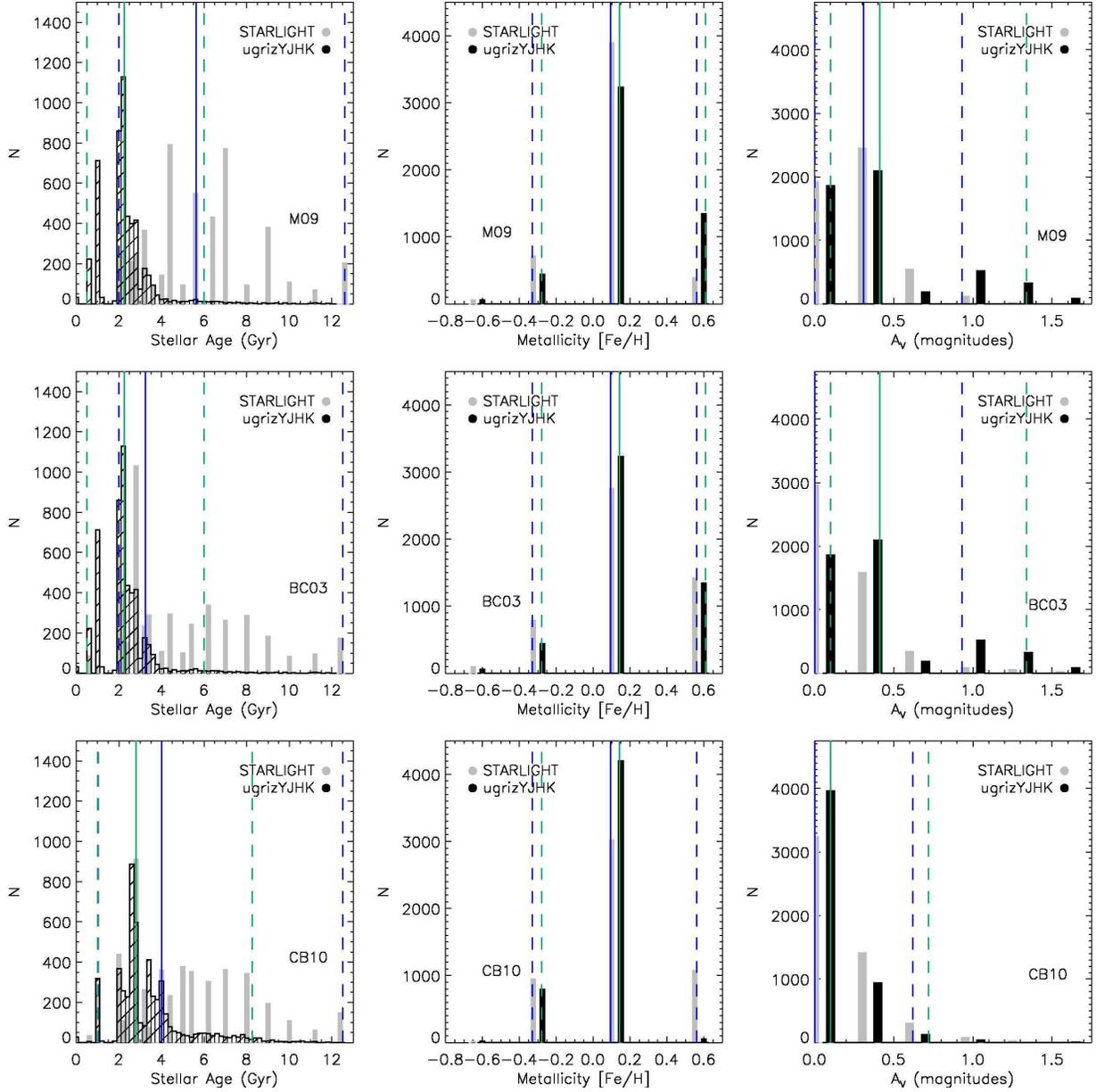}

\end{center}
\caption[XXX]
{\label{fig_starlight_v_lephare_all}
Comparison between spectroscopic ($gray$) and photometric ($black$) ages, metallicities, and extinctions for the optical+NIR sample. Stellar population models assume single bursts at $t=0$ and the parameters in Table \ref{tab_ssp}. {\it STARLIGHT} measured extinction has been binned to the six $E(B-V)$ values in the range $0-0.5$ (converted to extinction in magnitudes using $R_{V}=3.1$. $Solid$ $blue$ ($green$) lines show the sample median, measured with {\it STARLIGHT} ({\it LePhare}), with their 2$\sigma$ uncertainties shown as $dashed$ lines.
}
\end{figure}

\clearpage

\clearpage

\begin{multicols}{2}
\bibliography{arxiv}
\end{multicols}

\end{document}